\documentclass[11pt,a4paper]{article}
\pdfoutput=1

\usepackage[T1]{fontenc}
\usepackage{geometry}
\usepackage[authoryear,round]{natbib}

\usepackage{amsmath,amssymb,amsfonts}
\usepackage{graphicx}
\usepackage{booktabs}
\usepackage{authblk}
\usepackage{microtype}

\usepackage[colorlinks=true, linkcolor=blue, citecolor=blue, urlcolor=blue]{hyperref}

\DeclareRobustCommand{\VAN}[3]{#2}
\let\VANthebibliography\thebibliography
\def\thebibliography{\DeclareRobustCommand{\VAN}[3]{##3}\VANthebibliography}

\newcommand{\keywords}[1]{\noindent\textbf{Keywords:} #1\par\vspace{1em}}

\title{\textbf{A Machine Learning Approach to Finslerian Cosmological Dark Energy Model}}

\author[1]{Shruti Seth \thanks{Shruti Seth: sshruti.dstcims.2025@bhu.ac.in}}
\author[1]{R. Chaubey \thanks{R. Chaubey: rchaubey@bhu.ac.in}}

\affil[1]{Centre for Interdisciplinary Mathematical Sciences, Banaras Hindu University, Varanasi, Uttar Pradesh 221005, India}

\date{\today}

\begin{document}

\maketitle

\begin{abstract}

Modified extensions of the standard cosmological models often introduce complex parameter degeneracies that severely impact sampling convergence of the parameter space while analyzing the model observationally. In this work, we present an accelerated Bayesian inference technique using a Heteroscedastic Deep Ensemble with Monte Carlo Dropout incorporating distance-gated local residual corrections to emulate complex likelihood surfaces and avoid extrapolation beyond the trained region. We demonstrate this methodology on a unified dark energy cosmological model equipped with Finsler spacetime. The model is studied using the recent observational datasets, including Cosmic Chronometers (CC), Pantheon$+$SH$0$ES, and DESI DR$2$ BAO measurements. While the conventional sampling procedure struggles with slow mixing of parameter chains ($\tau_{max}=1578.3$), the neural network, specially trained densely in high likelihood regions, yields converged and highly resolved posterior distributions of the model parameters without significantly compromising the minimum chi-square value. We further discuss the results for physical inferences and also compare the present dark energy model with $\Lambda\text{CDM}$ statistically using AIC and BIC. 

\end{abstract}



\section{Introduction}
\label{sec:Intro}

The mystery of the accelerated expansion of the Universe and the nature of dark energy has been a prominent area of research over the years \citep{Riess1998, Perlmutter1999}. In this context, the standard cosmological model, $\Lambda\text{CDM}$, has achieved remarkable success in describing the large-scale structure and expansion history of the Universe \citep{Planck2020Cosmology}. However, it faces persistent observational and theoretical challenges, including the Hubble tension and the cosmological constant problem, while the possibility of departures from a constant late-time dark-energy component continues to motivate investigation beyond the standard framework. These issues have encouraged the exploration of alternative descriptions of gravitational dynamics and the underlying spacetime geometry. Among such approaches, Finsler geometry provides a generalization of Riemannian geometry in which the geometric structure can depend explicitly on the direction in the tangent space, offering a natural framework for investigating intrinsic anisotropy \citep{Chaubey2019Finsler}. Finslerian cosmological models have subsequently been explored in connection with dark-energy dynamics and late-time cosmic expansion \citep{Raushan2020Finsler, Angit2022, Singh2025FinslerRanders, Chakraborty2026FinslerRanders}. This direction-dependent geometric structure gives rise to non-trivial geometric contributions characterized by quantities such as the Cartan torsion tensor \citep{Asanov1985, Nekouee2025}. Finsler--Randers cosmological models lead to modified Friedmann and Raychaudhuri equations containing additional anisotropic geometric contributions, which can be investigated as possible alternatives to a cosmological constant in describing the late-time expansion of the Universe \citep{Stavrinos2005, Triantafyllopoulos2018, Nekouee2025, PRAVEEN2024}.

To complement this anisotropic geometric modification, one requires a suitable description of the matter and energy content of the Universe. Unified dark-sector models offer an approach in which dark matter and dark energy (coupled together as the mysterious dark content of the Universe) are described through a single effective fluid with a non-linear equation of state. The Chaplygin gas concept was introduced into cosmology through the standard Chaplygin gas, given as $p=-A/\rho$, which extends from a pressureless matter-like regime at early times ($p\approx0$) to a de Sitter-like regime at late times ($p\approx-\rho$) \citep{Kamenshchik2001}. This formulation was subsequently generalized to the generalized Chaplygin gas (GCG), $p=-A/\rho^\alpha$, introducing an additional parameter that provides greater flexibility in describing the cosmic expansion history \citep{Bento2002, Gorini2003}. Extending this framework, the Modified Chaplygin Gas (MCG) incorporated a linear barotropic term, $p=A\rho-B/\rho^\alpha$, allowing the fluid to accommodate early-time barotropic behaviour while retaining a negative pressure contribution at late times \citep{Benaoum2002MCG, Debnath2004}. The MCG framework has since been investigated through analytical studies and observational constraints, as well as in modified gravity extensions \citep{Lu2008MCG, Sharif2017MCG, Li2019MCG, Chaubey2009, Shukla2023_MCG, Chaubey2017, Chaubey2016_fRT, praveen2026finsler, NEKOUEE202419}. Incorporating the extended geometric framework, that is, Finsler--Randers, with the MCG description in cosmology has been discussed in this paper.

In contrast to purely theoretical studies, an observational approach provides validation for our investigation. In recent studies, this approach has been used to report the constraint values of the model parameters and to compare the newly developed model to the existing standard models through different statistical entities. Late-time geometric probes such as DESI DR$2$ BAO ($0.295 \le z \le 2.330$) \citep{Adame_2025, DESI2025DR2}, Type Ia Supernovae \citep{Scolnic2018Pantheon, PantheonPlus2022, Scolnic2022PantheonPlusData}, and Cosmic Chronometers (CC) \citep{Jimenez_2002, Moresco2016} allow to map out the universe's exact expansion history. The observational constraints on model parameters are crucial to understanding the behaviour and physical outcomes of extended frameworks, and these are obtained through the statistical sampling technique MCMC (Markov Chain Monte Carlo). The conventional algorithm used for this sampling technique in cosmology poses a great challenge in terms of computational time and sampling convergence in the case of non-standard dark-energy models. 

To overcome this, an effective machine-learning technique can provide an alternative to conventional sampling and is being used in practice by several authors \citep{Alsing2019, Wang2020_ECoPANN, Williams2021_Nessai, Arico2021_Emulators, SpurioMancini2022, Miller2022_Swyft, Wong2023_FlowMC, EscamillaRivera2020}. Among all, Heteroscedastic neural networks allow the predictive variance to vary with the input, providing a representation of input-dependent predictive uncertainty \citep{Kendall2017}. Monte Carlo (MC) dropout introduced a computationally simple way of estimating model uncertainty by retaining dropout during inference and interpreting the resulting stochastic predictions in an approximate Bayesian framework \citep{Gal2016Dropout}. Deep ensembles subsequently provided an alternative and complementary approach in which independently trained neural networks are combined to characterize epistemic uncertainty \citep{Lakshminarayanan2017}. In this paper, these approaches are combined within a neural surrogate framework to approximate the cosmological likelihood efficiently, with uncertainty estimates used to assess the reliability of the surrogate across parameter space.  

The key motivation of this paper is to resolve the computational and sampling bottleneck for a non-standard dark-energy cosmological model. For this, we firstly derive the dark-energy model theoretically, incorporating the Finslerian framework with MCG in section~\ref{sec:Mathematical_Framework}. Section~\ref{sec:Observational_Analysis} presents the conventional sampling of the model parameters, highlighting the bottleneck and establishing the need to adopt an efficient alternative approach. In section~\ref{sec:Ml_training}, we train the neural network (Heteroscedastic Deep Ensemble with MC Dropout) to learn the likelihood surface of the parameter space with respect to the joint observational datasets. Section~\ref{sec:ML_MCMC} again deals with parameter sampling, but this time with the predicted chi-square as a replacement for the exact chi-square in the same conventional sampling algorithm. It further highlights the differences and the reasons for the differences in the constraint value of each parameter obtained from both sampling techniques. Section~\ref{sec:Results} is based on the physical discussion of the constraint values along with the model comparison. Lastly, section~\ref{sec:Conclusion} presents the conclusion of the paper.

\section{Finsler-Randers Cosmological Model with Modified Chaplygin Gas (Finslerian--MCG)}
\label{sec:Mathematical_Framework}

\subsection{Finsler Geometry and the Randers Metric}
Unlike the standard Riemannian geometry, where the metric tensor $g_{\mu \nu}(x)$ depends only on the spacetime coordinates $x^\mu$, a Finsler space defines a fundamental length function $F(x,y)$ on the tangent bundle $TM$ which depends on both position $x^\mu$ and directional tangent vectors $y^\mu = \mathrm{d}x^\mu /\mathrm{d}\tau$ where $\tau$ is identified with proper time \citep{Asanov1985, Bao2000}. The length function $F(x,y)$ is smooth for $y\neq0$ and positively homogeneous of degree one in $y$. The fundamental Finsler metric tensor and Cartan torsion tensor are defined as 
\begin{align}
    f_{\mu \nu}(x,y) &= \frac{1}{2} \frac{\partial ^2F^2(x,y)}{\partial y^\mu \partial y^\nu}, \label{eq:Finsler_metric_def} \\
    C_{\mu \nu \lambda}(x,y) &= \frac{1}{2}\frac{\partial f_{\mu \nu}(x,y)}{\partial y^\lambda}. 
    \label{eq: cartan_torsion_def}
\end{align}
The Cartan torsion characterizes the directional dependence of the Finsler metric and vanishes identically when the geometry reduces to the Riemannian case \citep{Bao2000}.

The fundamental metric function in the Finsler-Randers geometric framework is given as 
\begin{equation}
    F(x,y)= \alpha(x,y) + \beta(x,y),
\end{equation}
where \citep{Bao2000}
\begin{equation}
    \alpha(x,y)=\sqrt{a_{\mu \nu}(x) y^\mu y^\nu}, \qquad \beta(x,y) = u_\mu(x)y^\mu.
\end{equation}
In the above, $a_{\mu \nu}(x)$ denotes the ordinary Riemannian metric and $u_\mu$ denotes the directional, anisotropic deformation of the geometry. The latter is introduced to represent a weak primordial field satisfying $|u_\mu| \ll 1$ \citep{Stavrinos2008, Chakraborty2026FinslerRanders}. The associated Randers fundamental tensor is 
\begin{equation}
    f_{\mu \nu}(x,y)= g_{\mu\nu} + \frac{1}{4\alpha}\left(u_\mu y_\nu + u_\nu y_\mu \right) - \frac{\beta}{\alpha^3}y_\mu y_\nu +u_\mu u_\nu, \qquad g_{\mu\nu}= \frac{F}{\alpha}a_{\mu \nu}(x).
\end{equation}
Since $F/\alpha=1+\beta/\alpha$, all deviations from the ordinary Riemannian metric are controlled by the Randers 1-form $u_\mu$, with the explicit $u_\mu u_\nu$ term existing at second order \citep{Stavrinos2008}.

\subsection{Mathematical Formulation and Basic Equations}
In this article we consider a spatially flat Robertson--Walker background,
\begin{equation}
    a_{\mu\nu} = \mathrm{diag}\left(1, -a^2(t), -a^2(t), -a^2(t) \right).
\end{equation}
Cosmological homogeneity and spatial isotropy restrict the Randers 1-form to the temporal direction, $u_\mu = (u_0(t),0,0,0)$. The corresponding Finsler quantities along the comoving four-velocity are evaluated as $y^\mu=(1,0,0,0)$. Within the weak-anisotropy approximation, the Cartan torsion is of order $(u)$ and is neglected at the order considered. Therefore, the corresponding osculating Riemannian metric is taken as
\begin{equation}
    r_{\mu\nu}(x) = f_{\mu\nu}\bigl(x,y(x)\bigr),
\end{equation}
whose connection is approximated by the ordinary Christoffel symbols constructed from $r_{\mu\nu}$ \citep{Stavrinos2008}.

From the weak-anisotropy Finsler--Randers construction adopted in \cite{Chakraborty2026FinslerRanders}, the generalized field equations are written as
\begin{equation}
    L_{\mu\nu}= 8 \pi G \left(T_{\mu \nu} - \frac{1}{2} T g_{\mu \nu} \right),
    \label{eq:finsler_einstein}
\end{equation}
where $L_{\mu \nu}$ denotes the Ricci tensor \citep{PRAVEEN2024} associated with the osculating metric. For a comoving perfect fluid,
\begin{equation}
    T_{\mu \nu}=(\rho + p)y_\mu y_\nu - p f_{\mu\nu}.
\end{equation}
Under the slow-variation approximations $\dot{u}_0^2 \simeq 0$ and $\ddot{u}_0 \simeq 0$, the modified Friedmann-like equations are given as \citep{Stavrinos2008, Chakraborty2026FinslerRanders}

\begin{align}
    \dot{H}+H^2+\frac{3}{4}H\dot{u}_0 &=-\frac{4\pi G}{3}(\rho+3p_{\text{MCG}}),
    \label{eq:f1} \\
    \dot{H}+3H^2+\frac{11}{4}H\dot{u}_0 &=4\pi G(\rho-p_{\text{MCG}}),
    \label{eq:f2}
\end{align}
where $H \equiv \dot{a}/a$.

\subsection{Finsler-Randers with MCG}
For the spatially flat case considered, from Eq.~\ref{eq:f1} and Eq.~\ref{eq:f2},
\begin{equation}
    3H^2=8\pi G(\rho+\rho_{F}), \qquad \rho_F \equiv -\frac{3H\dot{u}_0}{8\pi G}.
    \label{eq:friedmann_comb}
\end{equation}
Thus, the Finslerian correction can be represented at the background level by an effective density contribution $\rho_F$. For an accelerated expanding universe ($H>0$), $\rho_F>0$ corresponds to $\dot{u}_0<0$. The present work assumes the anisotropy to be linearly evolving, that is, $\dot{u}_0=\lambda H$. Therefore, the anisotropic contribution is directly characterized by $\lambda$ which must be negative ($\lambda < 0$) for an expanding universe ($H > 0$). The modified Raychaudhuri equation in a weak-anisotropy Robertson-Walker spacetime is expressed as
\begin{equation}
    \frac{d\Theta}{dt}+\frac{\Theta^2}{3}=-\frac{4\pi G}{3}(\rho+3p_{\text{MCG}})-\frac{3}{4}H\dot{u}_0, 
    \label{eq:raychaudhuri}
\end{equation}
where the expansion parameter $\Theta=3H$ and the modified energy-momentum conservation equation is given by
\begin{equation}
    \dot{\rho}+3H(\rho+p_{\text{MCG}})+\dot{u}_0\left(\frac{3}{2}p_{\text{MCG}}+\rho\right)=0.
    \label{eq:conservation}
\end{equation}

The Modified Chaplygin Gas equation of state as a unified description of dark energy is defined as \citep{Benaoum2002MCG, Debnath2004, Lu2008MCG}, 
\begin{equation}
    p_{\mathrm{MCG}} = A\rho - \frac{B}{\rho^\alpha},
    \label{eq:mcg_eos}
\end{equation}
where $A$, $B$ and $\alpha$ are model parameters. Taking this together with $\dot{u}_0=\lambda H$ in Eq.~\ref{eq:conservation}, the scale-factor-dependent energy density is obtained as
\begin{equation}
    \begin{split}
        \rho(a) = \Biggl[ & \frac{3\left(1+\frac{\lambda}{2}\right)}{3(1+A)+\lambda \left(1+\frac{3A}{2}\right)} B \\
        & + \left( \rho_0^{\alpha + 1} - \frac{3\left(1+\frac{\lambda}{2}\right)}{3(1+A)+\lambda \left(1+\frac{3A}{2}\right)} B \right) a^{-(1+\alpha)\left[3(1+A)+\lambda \left(1+\frac{3A}{2}\right)\right]} \Biggr]^{\frac{1}{1+\alpha}}
    \end{split}
    \label{eq:rho_a}
\end{equation}

The scale parameter and the redshift parameter are connected by $a=(1+z)^{-1}$. From Eq.~\ref{eq:rho_a} and Eq.~\ref{eq:friedmann_comb} the expansion rate is obtained as
\begin{equation}
    \begin{split}
        H(z)=\frac{1}{\sqrt{3(1+\lambda)}}\Biggl[\frac{3\left(1+\frac{\lambda}{2}\right)B}{3(1+A)+\lambda \left(1+\frac{3A}{2}\right)}+\left(\rho_0^{\alpha+1}-\frac{3\left(1+\frac{\lambda}{2}\right)B}{3(1+A)+\lambda \left(1+\frac{3A}{2}\right)}\right)\\
        (1+z)^{(1+\alpha)\left[3(1+A)+\lambda \left(1+\frac{3A}{2}\right)\right]} \Biggr]^{\frac{1}{2(1+\alpha)}}
    \end{split}
\end{equation}

Here, we define the dimensionless parameter 
\begin{equation}
   A_s = \frac{3\left(1+\frac{\lambda}{2}\right)B}{3(1+A)+\lambda \left(1+\frac{3A}{2}\right)\rho_0^{\alpha+1}},
   \label{eq:A_s}
\end{equation}
 as the effective dark energy fraction, which closely corresponds to $\Omega_{\Lambda}$ in the standard limit \citep{Benaoum2002MCG, Debnath2004, Lu2008MCG} and including the present-day Hubble parameter $H_0$ (at $z=0$), the background expansion history becomes
\begin{equation}
    H(z)=H_0 \left[ A_s + (1-A_s)(1+z)^{(1+\alpha)\left[3(1+A)+\lambda\left(1+\frac{3A}{2}\right)\right]} \right]^{\frac{1}{2(1+\alpha)}}.
     \label{eq:Hz_final}
\end{equation}
Now it is to be observed that in the isotropic limit ($\lambda \to 0$) the internal Finslerian anisotropy vanishes and Eq.~\ref{eq:Hz_final} reduces to the standard isotropic MCG expansion history \citep{Benaoum2002MCG, Debnath2004, Lu2008MCG}:
\begin{equation}
    H(z) = H_0 \left[ A_s + (1-A_s)(1+z)^{3(1+A)(1+\alpha)} \right]^{\frac{1}{2(1+\alpha)}}.
\end{equation} 
This demonstrates that the present form recovers the conventional Riemannian framework in the standard limit. While the barotropic limit ($B \to 0$) yields the reduced energy density $\rho(a) \propto (1+z)^{\left(3(1+A)+\lambda(1+\frac{3A}{2})\right)}$ (from Eq.~\ref{eq:rho_a}), demonstrating the consistency with the work of \cite{Chakraborty2026FinslerRanders}.

\section{Observational Datasets and Conventional Analysis}
\label{sec:Observational_Analysis}
In this section, we discuss the observational datasets used, their likelihood formulation and their significance in analyzing the present Finslerian--MCG cosmological model. It describes the numerical implementation and the joint computational strategy adopted for the analysis. Lastly, we execute the Markov Chain Monte Carlo (MCMC) posterior sampling analysis using \texttt{emcee} \citep{ForemanMackey2019emcee} and the posterior plots are obtained using \texttt{GetDist} \citep{Lewis_2025}. The parameter space taken for the observational analysis of the derived cosmological model (Eq.~\ref{eq:Hz_final}) is $\Theta_p = \{A_s, A, \alpha, \lambda, H_0, M\}$, including the absolute magnitude $M$ for the Pantheon$+$SH$0$ES compilation.

\subsection{Observational Datasets}

\subsubsection{The Cosmic Chronometer (CC) Dataset}
The compilation of $30$ Cosmic Chronometer (CC) data points spanning the redshift range $0.07 \le z \le 1.965$~\citep{Moresco2016, Vagnozzi_2021} has been used in the present analysis. The CC method provides a direct, largely model-independent measurement of the Hubble parameter $H(z)$ by measuring the differential age evolution ($\Delta t$) of passively evolving, massive galaxies as a function of redshift ($z$) through the differential relation:
\begin{equation}
    H(z) = -\frac{1}{1+z} \frac{\mathrm{d}z}{\mathrm{d}t} \, .
    \label{eq:cc_method}
\end{equation}
This dataset spans the transition epoch from matter-dominated deceleration to dark-energy-dominated acceleration. Each data point consists of the redshift $z_i$, the observed expansion rate $H_{\text{obs}}(z_i)$ in $\mathrm{km\,s^{-1}\,Mpc^{-1}}$, and its corresponding observational uncertainty $\sigma_{H(z), i}$. Model parameters $\Theta_p$ are constrained using the $\chi^2$ likelihood function~\citep{Moresco2016, Vagnozzi_2021}:
\begin{equation}
    \chi^2_{\text{CC}}(\Theta_p) = \sum_{i=1}^{30} \frac{\left[ H_{\text{obs}}(z_i) - H_{\text{model}}(z_i, \Theta_p) \right]^2}{\sigma_{H(z), i}^2} \, .
    \label{eq:cc_chi2}
\end{equation}

\subsubsection{The Pantheon+SH0ES Dataset}
The Pantheon+ Type Ia Supernova ($\text{SNe~Ia}$) compilation and the SH0ES (Supernovae and $H_0$ for the Equation of State of Dark Energy) Cepheid calibrations provide an anchor-calibrated local distance-ladder dataset \citep{Riess2022SH0ES, Scolnic2022PantheonPlusData}. Pantheon+ provides relative luminosity distance measurements across a wide redshift range, while SH0ES anchors these observations through absolute distance calibrations from Cepheid variable stars in nearby host galaxies. Together, they construct a continuous distance ladder capable of determining both the local expansion rate ($H_0$) and the dark energy evolution.

The dataset comprises $1701$ light-curve measurements corresponding to $1550$ distinct $\text{SNe~Ia}$ spanning $0.00122 \le z \le 2.26137$. The dataset consists of a data vector and a full covariance matrix:
\begin{enumerate}
    \item Data Vector: It contains observational parameters including heliocentric and CMB-frame redshifts ($z_{\text{HEL}}$, $z_{\text{CMB}}$, $z_{\text{HD}}$), corrected apparent magnitudes ($m_{B_{\text{corr}}}$), and Cepheid-based distance moduli ($\text{CEPH\_DIST}$). An indicator variable ($\text{IS\_CALIBRATOR}$) identifies the calibrator subset for galaxies hosting both Cepheids and $\text{SNe~Ia}$, connecting the anchor scale to the Hubble flow.
    \item Covariance Matrix ($C_{\text{PS}}$): A $1701 \times 1701$ symmetric matrix accounting for statistical measurement uncertainties along the diagonal and off-diagonal systematic errors, including zero-point calibration offsets, dust corrections, host galaxy correlations, and peculiar velocity flows.
\end{enumerate}

Model parameters $\Theta_p$ are constrained by evaluating the $\chi^2$ likelihood function using $C_{\text{PS}}^{-1}$:
\begin{equation}
    \chi^2_{\text{PS}}(\Theta_p) = \Delta Q^T C_{\text{PS}}^{-1} \Delta Q \, ,
    \label{eq:chi2_pantheon_shoes}
\end{equation}
where the residual vector elements $\Delta Q_i$ are defined as:
\begin{equation}
    \Delta Q_i = 
    \begin{cases} 
        m_{B_{\text{corr}}, i} - M - \text{CEPH\_DIST}_i & \text{if } \text{IS\_CALIBRATOR}_i = 1 \, , \\ 
        m_{B_{\text{corr}}, i} - M - \mu_{\text{model}}(z_i, \Theta_p) & \text{otherwise} \, .
    \end{cases}
    \label{eq:pantheon_residual}
\end{equation}
The theoretical distance modulus $\mu_{\text{model}}(z_i, \Theta_p)$ is calculated from the luminosity distance $d_L$:
\begin{equation}
    d_L(z_i, \Theta_p) = (1 + z_{\text{HEL}, i}) \int_{0}^{z_{\text{HD}, i}} \frac{c}{H_{\text{model}}(z', \Theta_p)} \, \mathrm{d}z' \, ,
    \label{eq:luminosity_distance}
\end{equation}
\begin{equation}
    \mu_{\text{model}}(z_i, \Theta_p) = 5 \log_{10} \left( \frac{d_L(z_i, \Theta_p)}{1\,\mathrm{Mpc}} \right) + 25 \, .
    \label{eq:distance_modulus}
\end{equation}
Here, $z_{\text{HD}}$ is used inside the integral to trace the cosmological expansion history, while $z_{\text{HEL}}$ accounts for the $(1+z)$ observational redshift scaling in the heliocentric frame.

To avoid peculiar velocity uncertainties and local bulk-flow effects, a lower redshift threshold of $z_{\text{HD}} > 0.01$ is applied exclusively to non-calibrator SNe. Calibrated Cepheid hosts retain direct distance measurements ($\text{CEPH\_DIST}$) regardless of redshift. Consequently, the filtered dataset consists of $1657$ data points and the corresponding $1657 \times 1657$ reduced covariance matrix \citep{Audren:2012wb, Brinckmann:2018cvx}.

\subsubsection{The DESI DR2 BAO Dataset}
Baryon Acoustic Oscillation (BAO) measurements from the Dark Energy Spectroscopic Instrument (DESI) Data Release 2 (DR2) \citep{DESI2025DR2, DESI2025DR2BAO} provide an efficient assessment of late-time accelerating expansion dynamics. DESI employs 5000 robotic focal-plane positioners to measure spectra across multiple tracer populations spanning $0.1 < z < 4.2$. BAO measurements provide an independent geometric probe of the cosmic expansion. The sound horizon at the baryon drag epoch, $r_d$, serves as a standard ruler calibrated by early-universe physics, taking a fiducial value $r_d = 147.05\,\text{Mpc}$ corresponding to a drag epoch at $z_d \approx 1060$ \citep{Planck2020Cosmology}. Depending on the target sample and line-of-sight orientation, cosmological models parameterized by $\Theta_p$ are constrained using three distance observables:
\begin{enumerate}
    \item Transverse Comoving Distance Ratio ($D_M/r_d$): Assuming spatial flatness, it measures expansion perpendicular to the line of sight, where $D_{M, \mathrm{model}}(z)$ is given as:
    \begin{equation}
        D_{M, \mathrm{model}}(z, \Theta_p) = \int_{0}^{z} \frac{c}{H_{\mathrm{model}}(z', \Theta_p)} \, \mathrm{d}z' \, .
        \label{eq:transverse_distance}
    \end{equation}
    
    \item Radial Distance Ratio ($D_H/r_d$): It measures the line-of-sight expansion rate using the Hubble distance $D_{H, \mathrm{model}}(z)$ and defined as:
    \begin{equation}
        D_{H, \mathrm{model}}(z, \Theta_p) = \frac{c}{H_{\mathrm{model}}(z, \Theta_p)} \, .
        \label{eq:radial_distance}
    \end{equation}
    
    \item Volume-Averaged Distance Ratio ($D_V/r_d$): This is primarily used at lower redshifts (such as the Bright Galaxy Sample) where sample volume limits anisotropic decomposition. The angle-averaged distance $D_{V, \mathrm{model}}(z)$ combines the transverse and radial ratios:
    \begin{equation}
        D_{V, \mathrm{model}}(z, \Theta_p) = \left[ z \, D_{M, \mathrm{model}}^2(z, \Theta_p) \, D_{H, \mathrm{model}}(z, \Theta_p) \right]^{1/3} \, .
        \label{eq:volume_distance}
    \end{equation}
\end{enumerate}

DESI constructs these observables using four primary tracer populations across distinct redshift regimes: the Bright Galaxy Sample ($\text{BGS}$, $z < 0.4$), Luminous Red Galaxies ($\text{LRGs}$, $0.4 < z < 1.0$), Emission Line Galaxies ($\text{ELGs}$, $0.6 < z < 1.6$), and Quasars ($\text{QSOs}$, $z > 1.6$), along with the Lyman-$\alpha$ forest absorption field at high redshifts \citep{DESI2025DR2BAO}.
The observational data vector $D_{\text{obs}}$ consists of 13 measured distance ratios ($D_M/r_d, D_H/r_d, D_V/r_d$) across these tracer bins, as summarized in Table~\ref{tab:desi_dr2_data_vector} \citep{DESI2025DR2}.

\begin{table}[htbp]
    \centering
        \caption{\textbf{Summary of DESI DR2 BAO measurements used in this analysis. Listed are effective redshifts ($z_{\text{DESI}}$), measured distance ratios ($D_M/r_d, D_H/r_d, D_V/r_d$), observational uncertainties ($e$), and correlation coefficients ($\gamma$).}} 
        \label{tab:desi_dr2_data_vector}
        \vspace{2mm}
        \begin{tabular}{|l|c|c|c|c|c|c|}
        \hline
        Tracer & Redshift ($z_{\text{DESI}}$) & Observable & Value & Error ($e$) & Correlation ($\gamma$) \\
        \hline
        LRG1 & $0.510$ & $D_{M,1} / r_d$ & $13.587$ & $0.169$ & $-0.475$ \\
             &         & $D_{H,1} / r_d$ & $21.863$ & $0.427$ & \\
        \hline
        LRG2 & $0.706$ & $D_{M,2} / r_d$ & $17.347$ & $0.180$ & $-0.423$ \\
             &         & $D_{H,2} / r_d$ & $19.458$ & $0.332$ & \\
        \hline
        LRG3 + ELG1 & $0.934$ & $D_{M,3} / r_d$ & $21.574$ & $0.153$ & $-0.425$ \\
                    &         & $D_{H,3} / r_d$ & $17.641$ & $0.193$ & \\
        \hline
        ELG2 & $1.321$ & $D_{M,4} / r_d$ & $27.605$ & $0.320$ & $-0.437$ \\
             &         & $D_{H,4} / r_d$ & $14.178$ & $0.217$ & \\
        \hline
        QSO & $1.484$ & $D_{M,5} / r_d$ & $30.519$ & $0.758$ & $-0.489$ \\
            &         & $D_{H,5} / r_d$ & $12.816$ & $0.513$ & \\
        \hline
        Ly$\alpha$ & $2.330$ & $D_{M,6} / r_d$ & $38.988$ & $0.531$ & $-0.431$ \\
                   &         & $D_{H,6} / r_d$ & $8.632$ & $0.101$ & \\
        \hline
        BGS & $0.295$ & $D_V / r_d$ & $7.944$ & $0.075$ & N/A \\           
        \hline
    \end{tabular}
\end{table}

The likelihood function is evaluated as:
\begin{equation}
    \chi^2_{\text{DESI}}(\Theta_p) = \Delta D^T C_{\text{DESI}}^{-1} \Delta D \, ,
    \label{eq:chi2_desi}
\end{equation}
where residual vector elements are defined as $\Delta D_i = D_{\text{obs}, i} - \left(\frac{D_{\text{model}, i}(z_i, \Theta_p)}{r_d}\right)$. Ordering the observational data vector as \[D_{\mathrm{obs}} = [D_{M,1\ldots6},\,D_{H,1\ldots6},\,D_V^{\mathrm{BGS}}]^T ,\] the $13\times13$ symmetric covariance matrix is constructed from the quoted measurement uncertainties and the six reported correlations between the corresponding $D_M/r_d$ and $D_H/r_d$ measurements. Its elements are

\begin{equation}
    c_{ij}=
    \begin{cases}
        e_i^2, & i=j,\\[4pt]
        e_i.e_{i+6}.\gamma_i,
        & j=i+6,\quad 1\le i\le 6,\\[4pt]
        c_{ji}, & i>j,\\[4pt]
        0, & \text{otherwise},
    \end{cases}
    \label{eq: desi_covariance}
\end{equation}
where $\gamma_i$ denotes the correlation coefficient between the corresponding $D_M/r_d$ and $D_H/r_d$ measurements in the $i$-th anisotropic redshift bin.

DESI provides complementary geometric constraints on the late-time expansion history and dark-energy dynamics, while offering a sensitive probe of departures from the standard $\Lambda\mathrm{CDM}$ model \citep{DESI2025DR2}.

\subsection{Numerical Implementation and Joint Computational Strategy}

Evaluation of background expansion observables across $1657$ Pantheon$+$SH$0$ES data points and $13$ DESI DR2 BAO distance measurements requires repeatedly computing the comoving distance integral:
\begin{equation}
    D_M(z, \Theta_p) = c \int_{0}^{z} \frac{\mathrm{d}z'}{H_{\mathrm{model}}(z', \Theta_p)} \, ,
    \label{eq:comoving_distance}
\end{equation}
which is required for both the Pantheon$+$SH$0$ES luminosity distance ($d_L = (1+z_{\mathrm{HEL}}) D_M(z_{\mathrm{HD}})$) and the DESI DR2 BAO distance ratios ($D_M/r_d$ and $D_V/r_d$).

To execute MCMC parameter estimation efficiently without running individual numerical calculations for thousands of data points at every step, we implement a grid-based vectorized computational pipeline inspired by fast sampling algorithms in cosmological inference codes like \texttt{MontePython} \citep{Audren:2012wb, Brinckmann:2018cvx, Blas2011CLASS, Lesgourgues2011CLASS}. The pipeline operates as follows:

\begin{enumerate}
    \item Redshift Grid Generation: At each step in the parameter space $\Theta_p = \{A_s, A, \alpha, \lambda,$ $ H_0, M\}$, the reciprocal expansion rate $c / H_{\mathrm{model}}(z, \Theta_p)$ is evaluated over a 1D grid of $3000$ points spanning $z \in [0, z_{\mathrm{max}}]$. Moreover, $z_{\mathrm{max}} = 1.01 \times \allowbreak \max(z_{\mathrm{DESI}}, z_{\mathrm{SNe}}) \approx 2.35$ is adopted to ensure $1\%$ safety margin above the highest-redshift tracer to avoid extrapolation beyond the interpolation domain \citep{Blas2011CLASS, Lesgourgues2011CLASS}.
    
    \item Cumulative Quadrature and Interpolation: The comoving distance array $D_M(z_k)$ on the grid is computed simultaneously using the composite trapezoidal rule:
    \begin{equation}
        D_M(z_k) = \sum_{i=1}^{k} \frac{c}{2} \left[ \frac{1}{H(z_i, \Theta_p)} + \frac{1}{H(z_{i-1}, \Theta_p)} \right] (z_i-z_{i-1}) \, .
    \end{equation}
    Linear interpolation is then applied to $D_M(z_k)$ to evaluate comoving distances for all $1657$ supernova host redshifts ($z_{\mathrm{HD}}$) and DESI tracer redshifts ($z_{\mathrm{DESI}}$) at once \citep{Blas2011CLASS, Lesgourgues2011CLASS}.
    
    \item Cholesky Whitening for Pantheon+SH0ES: To evaluate $\chi^2_{\mathrm{PS}}$ without numerically inverting the full $1657 \times 1657$ covariance matrix $C_{\mathrm{PS}}$ at every step, we precompute its lower Cholesky decomposition $C_{\mathrm{PS}} = L L^T$. The residual vector $\Delta Q$ is whitened by solving $Ly = \Delta Q$ through forward substitution using optimized triangular solvers \citep{Audren:2012wb, Brinckmann:2018cvx}, yielding:
    \begin{equation}
        \chi^2_{\mathrm{PS}}(\Theta_p) = y^Ty \, ,
    \end{equation}
    which significantly improves numerical stability and reduces matrix computational expense.
\end{enumerate}

The joint log-likelihood function evaluated by the MCMC sampler across all observational datasets is given by:
\begin{equation}
    \ln \mathcal{L}_{\mathrm{total}}(\Theta_p) = -\frac{1}{2} \left[ \chi^2_{\mathrm{CC}}(\Theta_p) + \chi^2_{\mathrm{DESI}}(\Theta_p) + \chi^2_{\mathrm{PS}}(\Theta_p) \right] \, .
    \label{eq:joint_likelihood}
\end{equation}
This grid-based acceleration algorithm yields a significant speedup in likelihood evaluation while preserving numerical accuracy across the sampled posterior distribution.

\subsection{Markov Chain Monte Carlo (MCMC) Sampling Analysis}
Finslerian--MCG model parameters are constrained using an affine-invariant Markov Chain Monte Carlo (MCMC) ensemble sampler implemented through the \texttt{emcee} Python package \citep{ForemanMackey2019emcee}. The sampled parameter vector is given by $\Theta_p = \{A_s, A, \alpha, \lambda, H_0, M\}$, where $\{A_s, A, \alpha\}$ parameterize the background MCG fluid dynamics, $\lambda$ is the Finslerian anisotropy parameter, $H_0$ is the present-day Hubble constant in $\mathrm{km\,s^{-1}\,Mpc^{-1}}$, and $M$ is the Pantheon$+$SH0ES SNe~Ia absolute magnitude nuisance parameter. The posterior probability density function $\mathcal{P}(\Theta_p \mid \mathrm{Data})$ is evaluated using Bayes' theorem:
\begin{equation}
    \ln \mathcal{P}(\Theta_p \mid \mathrm{Data}) = \ln \mathcal{L}_{\mathrm{total}}(\Theta_p) + \ln \mathcal{P}_{\mathrm{prior}}(\Theta_p) + \mathrm{C} \, ,
\end{equation}
where $\mathrm{C}$ is an arbitrary normalization constant. The total joint log-likelihood is evaluated (Eq.~\ref{eq:joint_likelihood}) using the vectorised numerical implementation described above. The prior range $\Omega_{\text{prior}}$ for the model parameters are taken as $A_s \in (0.60, 0.80)$, $A \in (-1.00, 1.00)$, $\alpha \in (-1.00, 1.00)$, $\lambda \in (-1.0, 0.5)$, $H_0 \in (60.0, 90.0) \mathrm{\,km\,s^{-1}\,Mpc^{-1}}$, and $M \in (-20.5, -18.0)$. 

Ensemble chains are initialized using $N_{\mathrm{walkers}} = 64$ parallel walkers centered around the fiducial point $\Theta_0 = \{0.706, 0.196, -0.324, -0.167, 69.0, -19.38\}$ and evolved for $25{,}000$ iterations per walker. Sampling convergence is studied using the Gelman--Rubin $\hat{R}$ statistic \citep{GelmanRubin1992} ($\hat{R} - 1 < 0.01$) and the integrated autocorrelation time \citep{Sokal1997, GoodmanWeare2010}. Initial burn-in iterations are discarded before extracting marginalized posterior credible intervals $1\sigma$ ($68.3\%$) and $2\sigma$ ($95.4\%$).
 
\begin{figure}[htbp]
    \centering
    \includegraphics[width=0.92\linewidth]{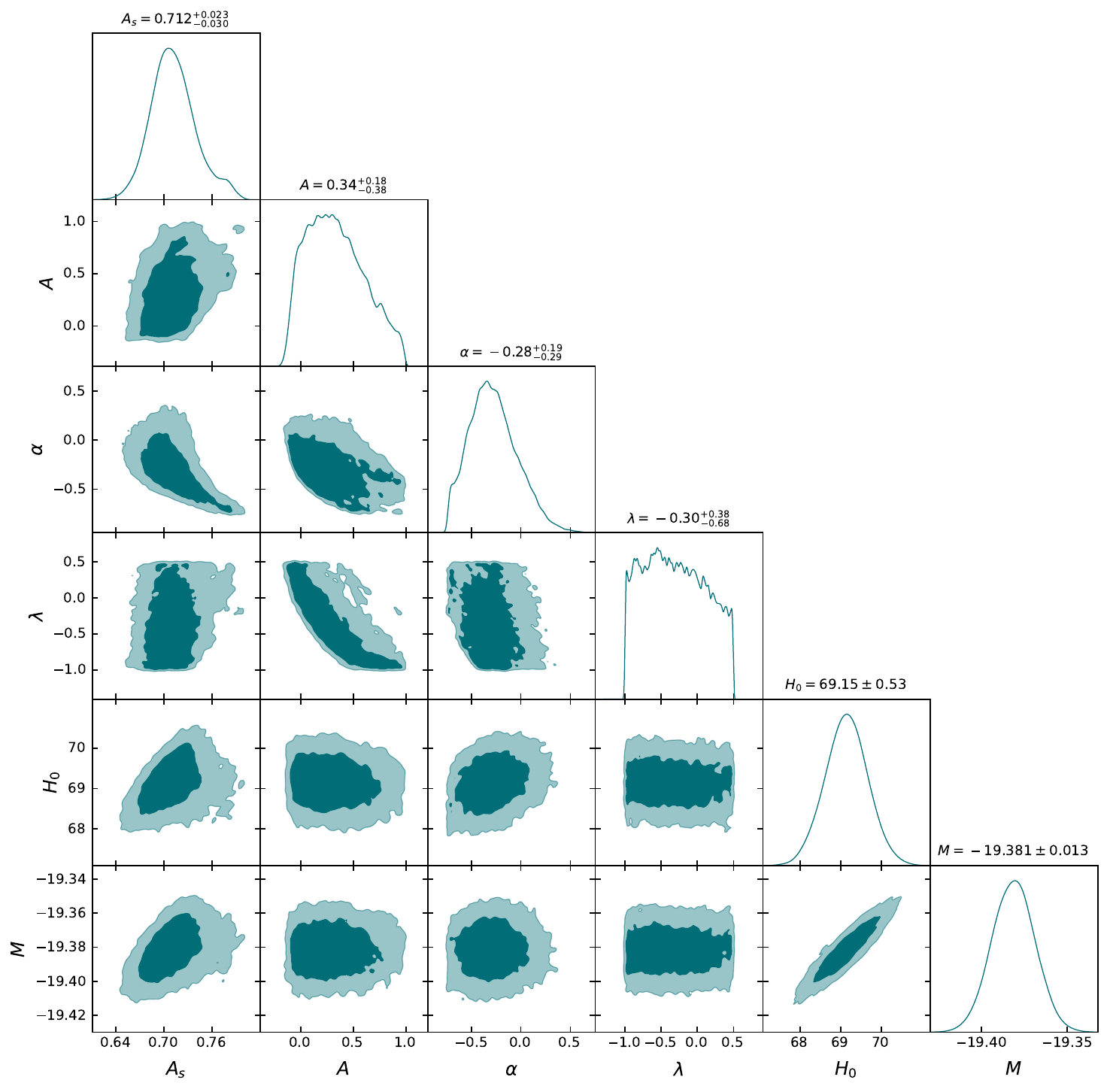}
    \caption{\textbf{Marginalized 1D posterior probability distributions and 2D joint posterior credible regions ($68.3\%$ and $95.4\%$ CL) for the Finslerian MCG model parameters obtained from the joint $\mathrm{CC} + \mathrm{Pantheon+SH0ES} + \mathrm{DESI\ DR2}$ dataset.}}
    \label{fig:corner_joint}
\end{figure}

The MCMC sampling yields 1D marginalized posterior distributions and 2D joint posterior credible regions ($68.3\%$ and $95.4\%$ CL), as shown in Figure~\ref{fig:corner_joint}. The joint likelihood optimization yields a minimum chi-square of $\chi^2_{\mathrm{min}} = 1501.40$. The extracted best-fit values, posterior statistics (mean, median, and asymmetric $68.3\%$ credible intervals), along with the MCMC convergence diagnostics ($\hat{R}$ and $\tau$), are summarized in Table~\ref{tab:mcmc_results}.

\begin{table}[htbp]
    \centering
    \caption{\textbf{Best-fit values, posterior estimates (Mean, Median ($68.3\%$ asymmetric credible intervals)), and convergence diagnostics ($\hat{R}$ and integrated autocorrelation time $\tau$) for the Finslerian MCG model parameters using the joint $\mathrm{CC + Pantheon+SH0ES + DESI\ DR2}$ dataset.}}
    \label{tab:mcmc_results}
    \vspace{2mm}
    \begin{tabular}{|l|c|c|c|c|c|c|}
        \hline
        Parameter & Best-fit & Mean & Median ($68.3\%$ Credible Interval) & $\hat{R}$ & $\tau$ (steps) \\
        \hline
        $A_s$ & $0.7126$ & $0.7119$ & $0.7099_{-0.0249}^{+0.0288}$ & $1.073$ & $1124.2$ \\[1.2ex]
        $A$ & $0.3966$ & $0.3357$ & $0.3098_{-0.2625}^{+0.3220}$ & $1.096$ & $1578.3$ \\[1.2ex]
        $\alpha$ & $-0.3604$ & $-0.2773$ & $-0.2976_{-0.2263}^{+0.2690}$ & $1.090$ & $1417.8$ \\[1.2ex]
        $\lambda$ & $-0.4635$ & $-0.2992$ & $-0.3208_{-0.4524}^{+0.5049}$ & $1.086$ & $1417.9$ \\[1.2ex]
        $H_0$ & $69.20$ & $69.15$ & $69.15_{-0.53}^{+0.52}$ & $1.021$ & $519.6$ \\[1.2ex]
        $M$ & $-19.379$ & $-19.381$ & $-19.381_{-0.013}^{+0.012}$ & $1.015$ & $449.6$ \\
        \hline
    \end{tabular}
\end{table}

 $H_0$ is inferred as  $69.15_{-0.53}^{+0.52}\,\mathrm{km\,s^{-1}\,Mpc^{-1}}$, exhibiting a nearly Gaussian $1$D posterior distribution. The SNe~Ia absolute-magnitude nuisance parameter is constrained to $M =-19.381_{-0.013}^{+0.012}$, making it one of the tightly constrained parameters in the analysis. The posterior distribution is observed to be nearly Gaussian and exhibits the smallest Gelman--Rubin statistic ($\hat{R}= 1.015$) along with the smallest autocorrelation length ($\tau=449.6$). Figure~\ref{fig:corner_joint} shows that $M$ exhibits a clear anti-correlation with $H_0$, which is a well-known feature of distance-ladder analysis, while Table~\ref{tab:mcmc_results} shows that the two parameters approach the ideal convergence threshold. 

The effective dark-energy parameter of the MCG framework is constrained to $A_s = 0.7099$ ${}^{+0.0288}_{-0.0249}$. Despite the posterior being nearly Gaussian, it exhibits a distorted asymmetric tail at the right end. The extended parameters $\{A,\alpha,\lambda\}$ exhibit elevated Gelman--Rubin statistics ($\hat{R}$) and long integrated autocorrelation length ($\tau$), indicating slow exploration of strongly correlated regions of parameter space within the given computational budget. Consequently, the inferred constraints on these parameters should be regarded as indicative rather than conclusive. The model parameter $A = 0.3098_{-0.2625}^{+0.3220}$ is weakly constrained. It exhibits the longest autocorrelation length ($\tau=1578.3$) and the highest Gelman--Rubin statistic ($\hat{R}=1.096$). The MCG exponent $\alpha$ is constrained to $-0.2976_{-0.2263}^{+0.2690}$ and remains weakly constrained owing to strong parameter degeneracies. The Finslerian anisotropy parameter $\lambda = -0.3208_{-0.4524}^{+0.5049}$ is poorly constrained, with its posterior distribution spanning a large fraction of the adopted prior range $(-1.0,0.5)$. As a result, the current posterior does not allow any conclusion about the physical interpretation of these parameters and the magnitude or significance of Finslerian anisotropy. The origin of this behaviour is physical. The effective background expansion exponent $(1 + \alpha) \left[ 3(1 + A) + \lambda\left(1 + \frac{3}{2}A\right) \right]$ allows simultaneous multi-parameter variations along extended degenerate manifolds in the $\{A, \alpha, \lambda\}$ subspace without significantly altering the background distance integrals $D_M(z)$ or expansion rate $H(z)$.

Therefore, the conventional MCMC analysis identifies the preferred region of the Finslerian--MCG parameter space and yields a minimum sampled chi-square, $\chi^2_{\mathrm{min}} = 1501.40$. However, the extended parameter sector $(A,\alpha,\lambda)$ exhibits pronounced degeneracies, resulting in long autocorrelation times ($\tau \approx (1.1-1.5)\times10^3$ steps) and Gelman--Rubin statistics that remain above the ideal convergence threshold. To achieve posterior convergence according to the adopted diagnostics, the sampling algorithm would require substantially longer chains and significantly increased computational cost. These results highlight a practical limitation of conventional Markovian sampling equipped with exact $\chi^2$ when applied to high-dimensional and highly correlated cosmological parameter spaces. This motivates the development of an alternative framework based on a machine-learning-assisted technique that can learn the likelihood surface from the sampled parameter space and provide a computationally efficient, uncertainty-aware approximation of the likelihood in the posterior-dominant region while substantially reducing the computational cost of repeated exact likelihood evaluations.

\section{Machine Learning Framework for Surrogate Likelihood Construction}
\label{sec:Ml_training}

To mitigate the high computational cost of conventional Bayesian sampling, we construct a machine-learning-assisted surrogate for the $\chi^2$ likelihood surface \citep{Alsing2019, SpurioMancini2022}, to replace the exact numerical evaluation while leaving the underlying Bayesian inference algorithm unchanged. The surrogate is a Heteroscedastic Deep Ensemble \citep{Nix1994, Lakshminarayanan2017} with Monte Carlo (MC) Dropout \citep{Gal2016Dropout}, incorporating a distance- and $\Delta\chi^2$- gated local residual corrector. The heteroscedastic output estimates input-dependent residual variance \citep{Kendall2017}, while the ensemble and MC-Dropout evaluations provide complementary epistemic-uncertainty assessments. These uncertainty estimates are used for surrogate validation and calibration only, while the \texttt{emcee} sampler  \citep{ForemanMackey2019emcee, GoodmanWeare2010} is executed by the deterministic (dropout-disabled) surrogate, ensuring reproducible likelihood evaluations. 

\subsection{Hybrid Training Dataset Construction}

The training candidate pool is designed to balance the broad prior-volume coverage with enhanced resolution around the Maximum A Posteriori (MAP) solution as,
\begin{equation}
    \mathcal{D}_{\mathrm{cand}} = \mathcal{D}_{\mathrm{LHS}} \cup \mathcal{D}_{\mathrm{local}} \cup \mathcal{D}_{\mathrm{ultra}}
\end{equation}
where,
\begin{align}
    \mathcal{D}_{\mathrm{LHS}} &\sim \mathrm{LHS}(\Omega_{\mathrm{prior}}), \\
    \mathcal{D}_{\mathrm{local}} &\sim \mathcal{N}(\Theta_{\mathrm{MAP}}, \mathrm{diag}(\sigma_{\mathrm{local}}^2)),\\
    \mathcal{D}_{\mathrm{ultra}} &\sim \mathcal{N}(\Theta_{\mathrm{MAP}},\mathrm{diag}(\sigma_{\mathrm{ultra}}^2))
\end{align}
where $\mathrm{LHS}(\Omega_{\mathrm{prior}})$ denotes Latin Hypercube Sampling \citep{McKay1979} of the prior volume, which provides near-uniform coverage of the six-dimensional parameter space at optimized sample cost, and
\begin{align}
    \sigma_{\mathrm{local}}&=
    \left[0.10,\,0.0020,\,0.06,\,0.06,\,1.0,\,0.05\right]^{T},\\
    \sigma_{\mathrm{ultra}}&=\left[0.02,\,0.0005,\,0.015,\,0.015,\,0.30,\,0.015\right]^{T}.
\end{align}
Candidates are filtered by the prior and physical constraints, after which the exact joint $\chi^2$ is evaluated for each retained point. The final training set $\mathcal{D}$ consisting of $1.5 \times 10^5$ points is formed by retaining the $80\%$ lowest-$\chi^2$ candidates for dense likelihood coverage, with the remaining $20\%$ from the rest of the pool to preserve broader coverage. This construction concentrates training density where posterior inference occurs while retaining global coverage for the prior space.

\subsection{Heteroscedastic Deep-Ensemble and MC-Dropout Surrogate}

The coarse surrogate maps the parameter vector $\Theta \in \Theta_p$ to a standardized regression target, 
\begin{equation}
    y(\Theta)=\frac{\log\left[1+\Delta\chi^2(\Theta)\right]-y_{\mathrm{mean}}}{y_{\mathrm{std}}}, \qquad \Delta\chi^2(\Theta)=\chi^2_{\mathrm{exact}}(\Theta)-\chi^2_{\mathrm{min}}
\end{equation}
with $y_{\mathrm{mean}}$, $y_{\mathrm{std}}$, and the input standardization computed from the training set. The $\log(1+\Delta\chi^2)$ transform compresses the long tail of the target distribution, preventing high-$\Delta\chi^2$ outliers from dominating the loss.

\begin{figure}[htbp]
    \centering
    \includegraphics[width=\textwidth]{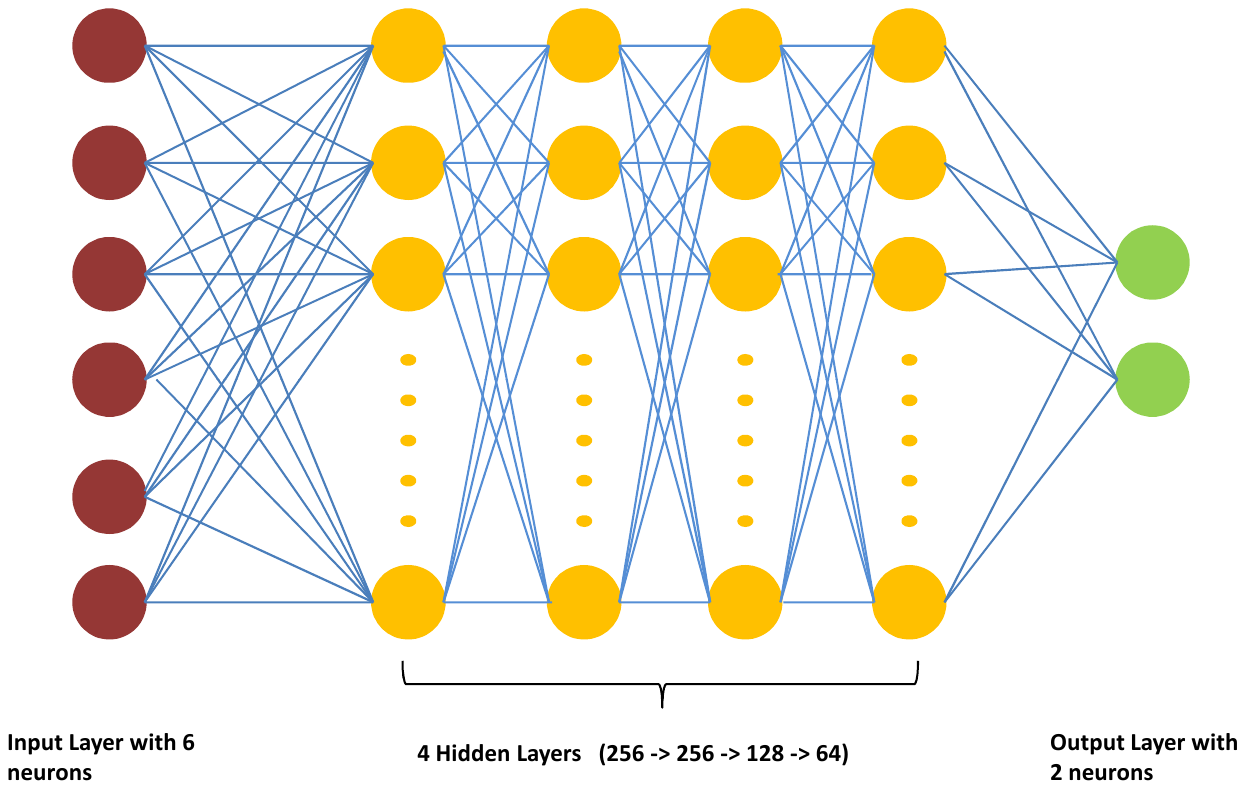}
    \caption{\textbf{Neural network architecture: $6$ inputs, $4$ hidden layers, $2$ outputs}}
    \label{fig:Network_Architecture}
\end{figure}

The network is a fully connected feed-forward neural-network architecture (Fig.~\ref{fig:Network_Architecture}) implemented using \texttt{PyTorch} \citep{Paszke2019} with SiLU activations \citep{Elfwing2018} and MC-Dropout ($p_{\mathrm{drop}}=0.08$) after each hidden layer,   
\begin{equation}
    6 \rightarrow 256 \rightarrow 256 \rightarrow 128 \rightarrow 64 \rightarrow (\hat{\mu},\hat{\ell}), \qquad \operatorname{SiLU}(x)=x\,\sigma(x), \ \ \sigma(x)=\frac{1}{1+\exp(-x)},
\end{equation}
where $\hat{\mu}$ is the predicted mean of the normalized target and $\hat{\ell}=\log\hat\sigma^2$ the predicted log-variance of the residual.

The network is trained by minimizing the Heteroscedastic Gaussian Negative Log-Likelihood (NLL) \citep{Nix1994, Kendall2017},
\begin{equation}
    \mathcal{L}_{\mathrm{NLL}}=\frac{1}{2N}\sum_{i=1}^{N}w_i\left[\exp(-\hat{\ell}_i)\left(y_i-\hat{\mu}_i\right)^2+\hat{\ell}_i\right], \qquad w_i \propto \left(1+\Delta\chi_i^2\right)^{-1/2},
\label{eq:heteroscedastic_nll}
\end{equation}
with weights normalized to unit mean. This amplifies the low $\Delta\chi^2$ region relevant to posterior inference without discarding the broader training domain. Five independently initialized realizations are trained with \texttt{AdamW} \citep{Loshchilov2019}, a cosine learning-rate schedule  \citep{Loshchilov2017SGDR}, gradient clipping, and early stopping on validation NLL. The deterministic point estimate is the dropout-disabled ensemble mean, while MC-Dropout is activated later with $40$ stochastic passes per ensemble member to estimate predictive uncertainty from the combined spread of the stochastic realizations and the heteroscedastic variance head \citep{Kendall2017}.

\begin{figure}[htbp]
    \centering
    \includegraphics[width=\textwidth]{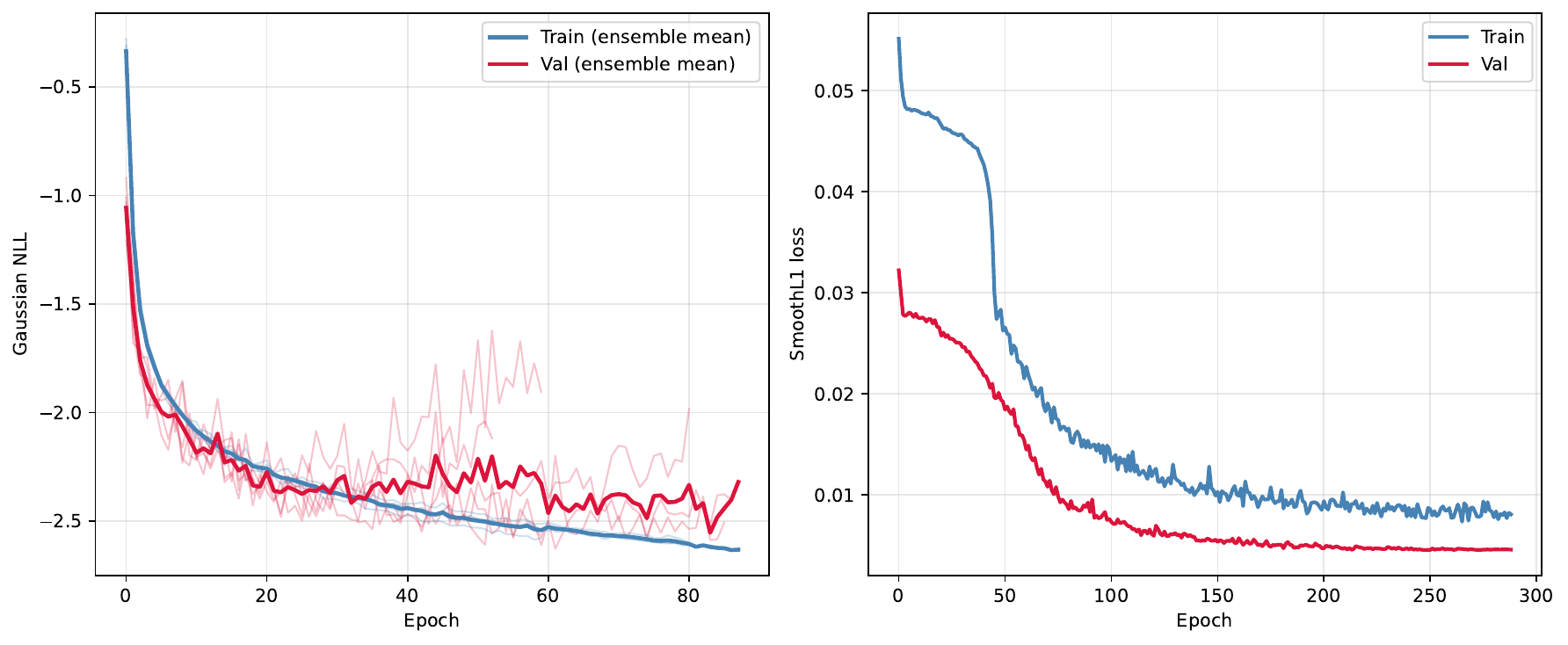}
    \caption{\textbf{Training and validation loss trajectories for the two-stage surrogate model.}
    \textit{Left:} Gaussian negative log-likelihood (NLL) loss curves during training of the coarse heteroscedastic deep ensemble. The individual ensemble-member trajectories are shown with reduced opacity, while the ensemble-mean training and validation losses are shown prominently. 
    \textit{Right:} Smooth $L_1$ loss convergence for the local residual corrector, showing the corresponding training and validation trajectories during its optimization.}
    \label{fig:surrogate_training_loss}
\end{figure}

\subsection{Local Residual Correction, Distance Gating, and Validation}

A local residual-correction network, trained only on points with $\Delta\chi^2 < 100$ is used to refine the coarse surrogate in the high-likelihood region,
\begin{equation}
    6 \rightarrow 128 \rightarrow 128 \rightarrow 64 \rightarrow 1,
\end{equation}
with SiLU activations and a dropout probability of $0.05$ in the first two hidden layers, and target
\begin{equation}
    r(\Theta) = \Delta\chi^2_{\mathrm{exact}}(\Theta)-\Delta\chi^2_{\mathrm{coarse}}(\Theta)
    \label{eq:local_residual}
\end{equation}
It is optimized with the Smooth $L_1$ (Huber) loss \citep{Girshick2015} ($\beta=0.3$), 
\begin{equation}
    \mathcal{L}_{\mathrm{Smooth}\,L_1}(e)=
    \begin{cases}
        \dfrac{e^2}{2\beta}, & |e|<\beta\\[3pt]
        |e|-\dfrac{\beta}{2}, & |e|\geq\beta,
        \end{cases}
    \label{eq:smooth_l1}
\end{equation}
This is used to limit the influence of outlier training points while retaining strong gradient signal near convergence. \texttt{AdamW} optimization with cosine scheduling and early stopping is used same as above.

To confine the correction to its validated domain, a $\texttt{cKDTree}$ \citep{Bentley1975, Virtanen2020} is built from the standardized local training points, giving a nearest-neighbour distance $d(\Theta)$ and out-of-domain threshold $d_{\mathrm{OOD}}=0.1060$ ($95^\mathrm{th}$ percentile of the reference nearest-neighbour distances). The two multiplicative gates jointly control the correction,

\begin{align}
    g_{\mathrm{dist}}(\Theta) &= \operatorname{clip}\left(1-\frac{d(\Theta)}{d_{\mathrm{OOD}}},0,1\right), \\
    g_{\chi^2}(\Theta) &= \frac{1}{1+\left[\operatorname{clip}\left(\frac{\Delta\chi^2_{\mathrm{coarse}}}
    {2\Delta\chi^2_{\mathrm{local}}},0,50\right)\right]^4}, \qquad
    \Delta\chi^2_{\mathrm{local}}=100, \\
    g(\Theta) &= g_{\mathrm{dist}}(\Theta)\,g_{\chi^2}(\Theta),
\label{eq:total_correction_gate}
\end{align}

where $\operatorname{clip}(x, a, b) \equiv \max(a, \min(x, b))$ bounds $x$ to the interval $[a, b]$. Here, $g_{\mathrm{dist}}(\Theta)$ suppresses out-of-distribution points beyond $d_{\mathrm{OOD}}$, while $g_{\chi^2}(\Theta)$ restricts corrections to the validated likelihood region. Combined, $g(\Theta)$ ensures the gate operates exclusively within the reliable surrogate manifold. The predicted residual is rescaled and clamped, $r_{\mathrm{corr}}= \operatorname{clip}(\hat r, -8s_r, 8s_r)$ with $s_r$ being the local residual standard deviation, and the corrected surrogate is 
\begin{equation}
    \Delta\chi^2_{\mathrm{corr}}(\Theta)=\max\left[0,\,\Delta\chi^2_{\mathrm{coarse}}(\Theta)+
    g(\Theta)r_{\mathrm{corr}}(\Theta)\right],
\label{eq:local_corrected_dchi2}
\end{equation}
which enforces $\Delta\chi^2 \geq0$ and prevents uncontrolled extrapolation of the correction. This gated, deterministic $\Delta\chi^2_{\mathrm{corr}}$ is the likelihood surface used by the MCMC sampler. 

The predictive reliability is further assessed using the absolute pull \citep{Dawid1982, Kuleshov2018},
\begin{equation}
    |z|=\frac{|\chi^2_{\mathrm{exact}}-\hat{\chi}^2|}{\sigma_{\mathrm{pred}}},
\end{equation}
calibrated with the RMS scale factor,
\begin{equation}
    s_{\mathrm{cal}}=\left[\frac{1}{N}\sum_i\left( \frac{\chi^2_{\mathrm{exact},i}-\hat{\chi}^2_i}{\sigma_{\mathrm{pred},i}}\right)^2\right]^{1/2}.
    \label{eq:calibration}
\end{equation}
Before calibration, the network's heteroscedastic variance head substantially overestimates its own predictive uncertainty, yielding the raw pull distribution median as $|z|=0.049$ ($90^\mathrm{th}$ percentile $0.151$), which is far below the expected value for a well-calibrated Gaussian residual ($\approx 0.674$). Therefore, applying the global RMS rescaling (Eq~\ref{eq:calibration}) yields  $|z|=0.536$ ($90^\mathrm{th}$ percentile $1.658$). The latter is substantially improved with respect to the ideal value and demonstrates that the uncertainty estimates, once rescaled, are neither over- nor under-confident in the posterior relevant region.
\begin{figure}[htbp]
\centering
    \includegraphics[width=\textwidth]{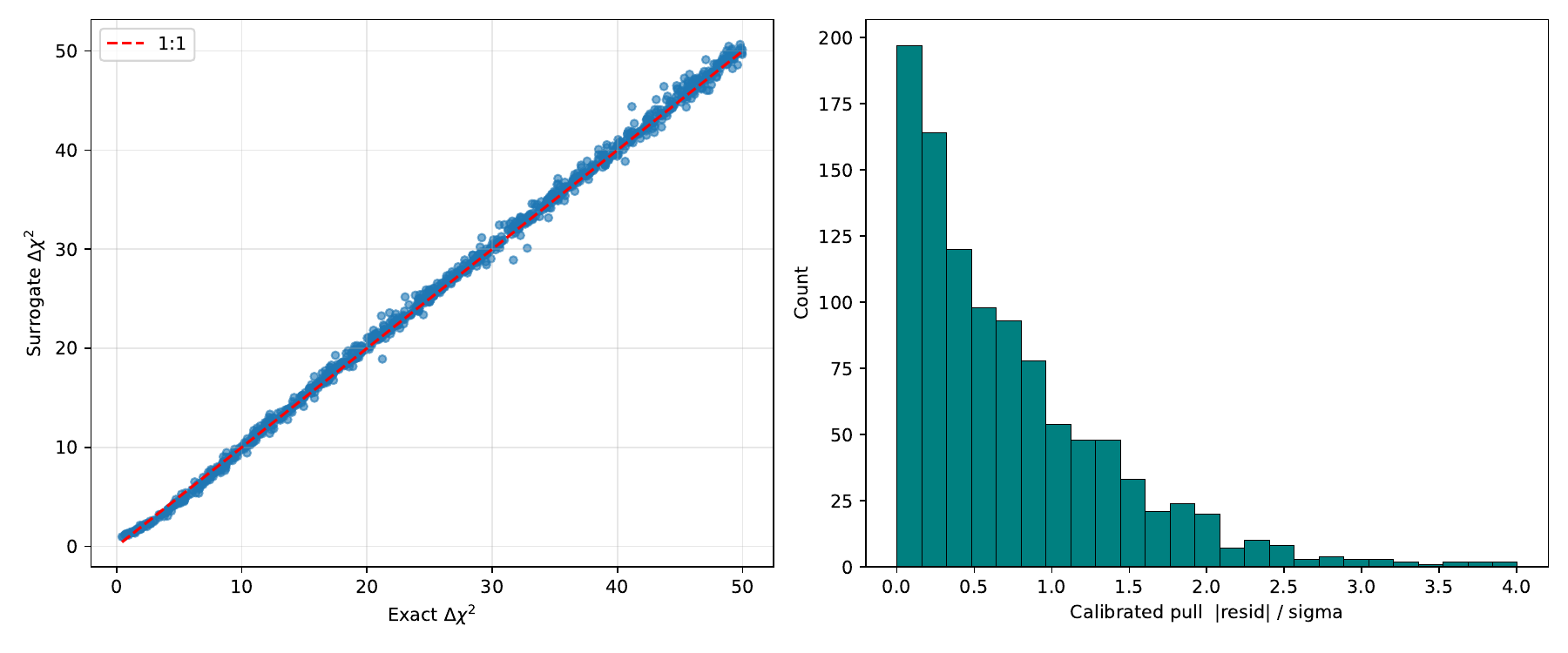}
    \caption{\textbf{Surrogate predictive performance and uncertainty calibration on the held-out validation set.}
    \textit{Left:} Comparison between the exact $\Delta\chi^2$ values and the deterministic surrogate predictions in the posterior-relevant
    near-minimum region. The red dashed line denotes the ideal $1{:}1$ relation, with the surrogate achieving $R^2=0.9983$.
    \textit{Right:} Distribution of the calibrated absolute predictive pull, providing a diagnostic of surrogate uncertainty calibration.}
    \label{fig:surrogate_performance_calibration}
\end{figure}

On the held-out validation set, the full corrected surrogate achieves a Mean Absolute Error $\mathrm{MAE} = 0.4170$, a Root Mean Square Error $\mathrm{RMSE} = 0.5770$ and a coefficient of determination $R^2 = 0.9983$ (Table~\ref{tab:surrogate_diagnostics}), confirming that the surrogate is ready to use within the adopted posterior-sampling domain. The subsequent MCMC analysis therefore replaces the exact $\chi^2$ evaluation with this validated surrogate, substantially reducing the computational cost of likelihood evaluation while preserving the original sampling procedure. 

\begin{table}[htbp]
    \centering
    \caption{\textbf{Key diagnostic metrics for the Heteroscedastic Deep Ensemble surrogate and local residual corrector.}}
    \label{tab:surrogate_diagnostics}
    \vspace{2mm}
    \begin{tabular}{|l|l|c|}
        \hline
        \textbf{Category} & \textbf{Metric} & \textbf{Value} \\
        \hline
        Heteroscedastic Deep Ensemble 
        & Number of ensemble members & $5$ \\
        & Coarse Near-Minimum MAE ($\Delta\chi^2$) & $0.6008$ \\
        & Coarse Near-Minimum RMSE ($\Delta\chi^2$) & $0.8087$ \\
        \hline
        Local Residual Corrector
        & Local Training Threshold & $\Delta\chi^2<100$ \\
        & OOD Distance Threshold & $0.1060$ \\
        & Maximum Correction Clamp ($\Delta\chi^2_{\mathrm{corr}}$) & $52.428$ \\
        \hline
        Complete Corrected Surrogate
        & Held-Out MAE ($\Delta\chi^2$) & $0.4170$ \\
        & Held-Out RMSE ($\Delta\chi^2$) & $0.5770$ \\
        & Held-Out $R^2$ & $0.9983$ \\
        & Calibrated Median $|z|$ & $0.536$ \\
        & Calibrated $90^{\mathrm{th}}$ Percentile $|z|$ & $1.658$ \\
        \hline
    \end{tabular}
\end{table}

\section{MCMC Sampling using ML-Surrogate Likelihood}
\label{sec:ML_MCMC}

In this section, we proceed with the execution of the MCMC sampling algorithm incorporating the above gated, deterministic likelihood.  The ensemble chains are initialized with $64$ walkers centered on $\Theta_0$ and evolved for $25,000$ iterations per walker. Convergence is further assessed for all six model parameters using the Gelman--Rubin diagnostic, with $\hat{R}<1.01$ \citep{GelmanRubin1992}. We further compare the posterior distributions obtained from both analyses.

To quantify the statistical difference between any two marginalized posterior distributions $\mathcal{P}_1(\theta)$ and $\mathcal{P}_2(\theta)$, we evaluate the parameter shift in units of the combined effective standard deviation:
\begin{equation}
    D_\sigma(\theta_1, \theta_2) = \frac{|\bar{\theta}_1 - \bar{\theta}_2|}{\sqrt{\sigma_{\mathrm{eff}, 1}^2 + \sigma_{\mathrm{eff}, 2}^2}}, \qquad  \quad \sigma_{\mathrm{eff}} = \frac{\sigma_+ + \sigma_-}{2}.
    \label{eq:unified_shift_metric}
\end{equation}
where, $\bar{\theta}$ denotes the marginalized posterior median, and $\sigma_{\mathrm{eff}}$ represents the symmetrized effective $1\sigma$ uncertainty derived from the upper ($\sigma_+$) and lower ($\sigma_-$) bounds of the $68.3\%$ credible intervals. 

\begin{figure}[htbp]
    \centering
    \includegraphics[width=\textwidth]{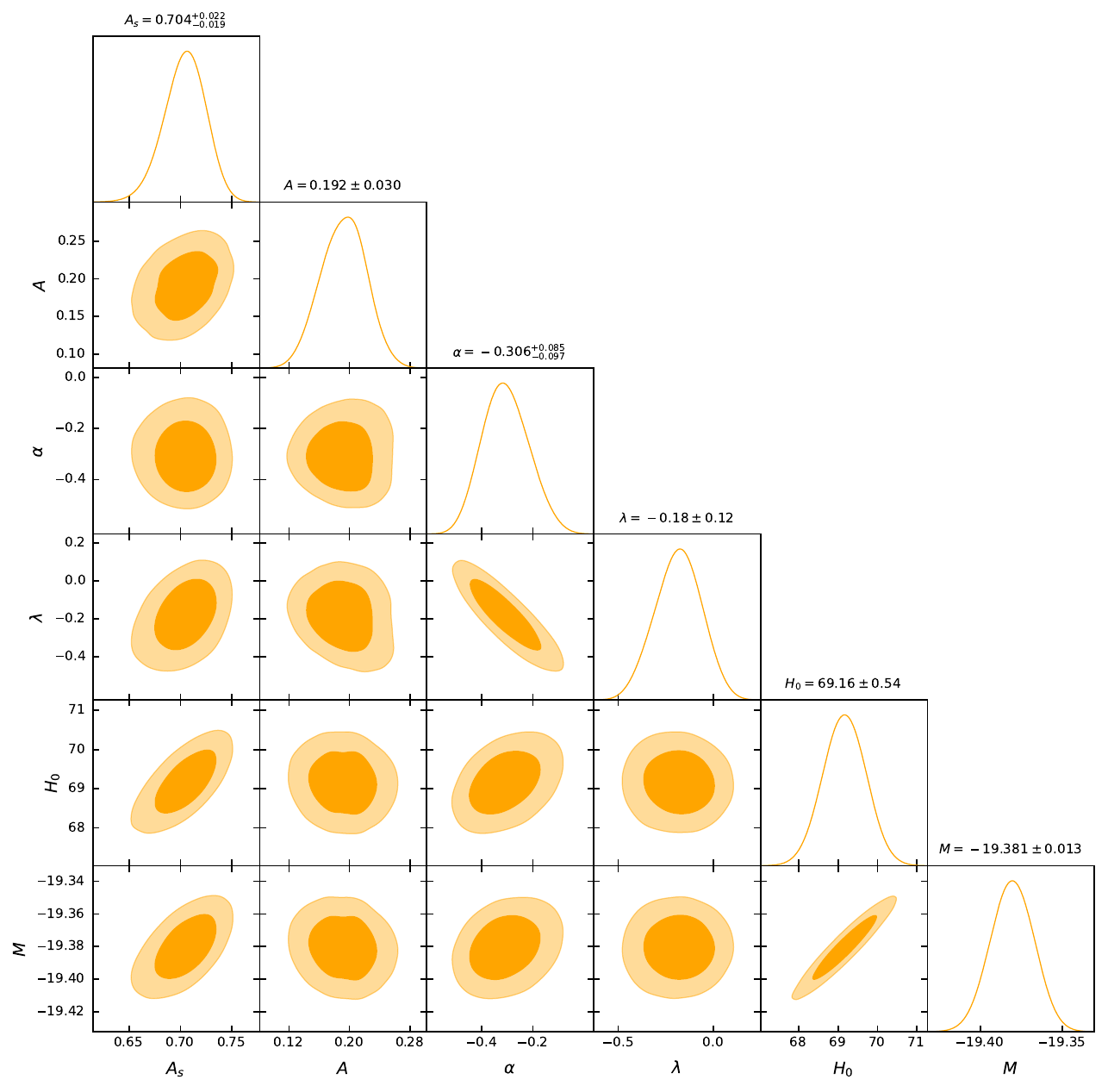}
    \caption{\textbf{Marginalized one-dimensional posterior distributions and two-dimensional joint posterior credible regions ($68.3\%$ and $95.4\%$) obtained from the surrogate-assisted MCMC analysis.}}
    \label{fig:bnn_triangle}
\end{figure}

Figure~\ref{fig:bnn_triangle} represents the marginalized one-dimensional posterior distributions and the corresponding two-dimensional joint posterior distributions obtained from the surrogate-assisted MCMC sampling. The contours enclose the $68.3\%$ and $95.4\%$ Bayesian credible regions. The resulting posterior distributions are smooth and well-defined, concentrated in the region of high likelihood. The surrogate is intentionally optimized for accurate representation of the high-likelihood region rather than for reproducing the poorly constrained low-probability tails of the exact-likelihood posterior. Consequently, some contraction of the marginalized posterior widths is expected for parameters whose exact chains exhibit broad tails and incomplete mixing.

\subsection{Surrogate-Assisted MCMC Constraints}

For the joint likelihood study, the conventional exact-likelihood MCMC analysis gives a minimum $\chi^2_{\min}=1501.40$. Using the trained neural-network surrogate, the production MCMC sampling recovers the same high-likelihood region, with the surrogate-derived best-fit point yielding a predicted $\chi^2_{\mathrm{Surr}}=1501.5215 \pm 0.2052$ with MC Dropout activated for predictive uncertainty estimation. When the corresponding best-fit parameter vector is reevaluated using the full exact likelihood, it gives $\chi^2_{\mathrm{Surr, exact}}=1501.3998$. Thus, the exact likelihood at the surrogate-derived solution differs from the minimum obtained in the conventional exact-likelihood MCMC analysis by only $\Delta\chi^2=0.0002$, while the absolute difference between the surrogate prediction and its exact reevaluation is $0.1217$. These small differences demonstrate that the surrogate-assisted sampler identifies essentially the same region as the computationally expensive exact-likelihood analysis but is more concentrated and highly resolved in the high-likelihood region, while substantially reducing the cost of repeated likelihood evaluations. The surrogate-assisted analysis yields the best-fit values for the model parameters $\{A_s, A, \alpha, \lambda,$ $H_0, M\}$ as $\{0.7081, 0.2166, -0.3371, -0.1937, 69.15, -19.380\}$. The extracted best-fit values, posterior statistics and their respective convergence diagnostics are summarized in Table~\ref{tab:ML_mcmc_results}

\begin{table}[htbp]
    \centering
    \caption{\textbf{Best-fit values, posterior estimates (Mean, Median ($68.3\%$ asymmetric credible intervals)), and convergence diagnostics ($\hat{R}$ and integrated autocorrelation time $\tau$) from the surrogate-assisted MCMC analysis.}}
    \label{tab:ML_mcmc_results}
    \vspace{2mm}
    \begin{tabular}{|l|c|c|c|c|c|c|}
        \hline
        Parameter & Best-fit & Mean & Median ($68.3\%$ Credible Interval) & $\hat{R}$ & $\tau$ (steps) \\
        \hline
        $A_s$     & $0.7081$  & $0.7042$  & $0.7050_{-0.0210}^{+0.0193}$  & $1.002$ & $85.9$  \\[1.2ex]
        $A$       & $0.2166$  & $0.1921$  & $0.1932_{-0.0317}^{+0.0280}$  & $1.002$ & $84.4$  \\[1.2ex]
        $\alpha$  & $-0.3371$ & $-0.3059$ & $-0.3096_{-0.0876}^{+0.0963}$ & $1.002$ & $108.1$ \\[1.2ex]
        $\lambda$ & $-0.1937$ & $-0.1839$ & $-0.1813_{-0.1276}^{+0.1205}$ & $1.003$ & $111.9$ \\[1.2ex]
        $H_0$     & $69.15$   & $69.16$   & $69.16_{-0.54}^{+0.54}$       & $1.002$ & $82.6$  \\[1.2ex]
        $M$     & $-19.380$ & $-19.381$ & $-19.381_{-0.013}^{+0.013}$   & $1.002$ & $83.2$  \\
        \hline
    \end{tabular}
\end{table}

The convergence diagnostics reported in Table~\ref{tab:ML_mcmc_results} indicate stable sampling for all six parameters. The Gelman--Rubin statistic satisfies $\hat{R}<1.01$ for every parameter, while the integrated autocorrelation times are substantially smaller, indicating that the chains contain many correlation lengths and provide adequate sampling for posterior estimation.

\subsection{Comparison with the Exact-Likelihood Posterior}

\begin{figure}[htbp]
\centering
    \includegraphics[width=\textwidth]{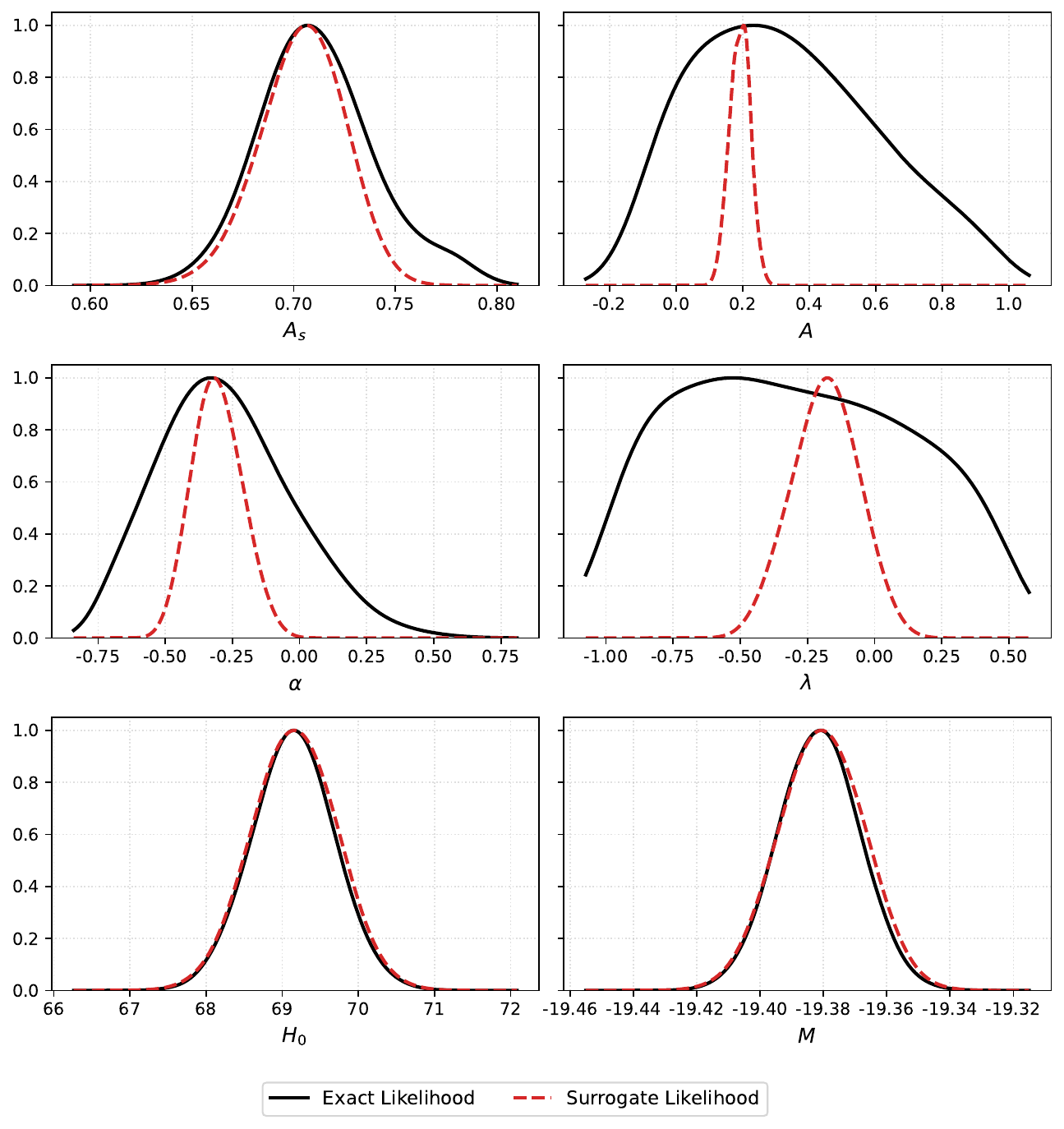}
    \caption{\textbf{Comparison of normalized one-dimensional marginalized posterior probability density functions (PDFs) obtained from the exact likelihood (black solid line) and the surrogate likelihood (red dashed line).} The baseline expansion parameters $H_0$ and $M$ exhibit near-perfect agreement, while parameters $A$, $\alpha$, and $\lambda$ show increased posterior concentration under surrogate sampling, reflecting convergence within the high-likelihood region compared to the unconverged exact chains.}
\label{fig:posterior_overlay_1D}
\end{figure}

\begin{figure}[htbp]
\centering
    \includegraphics[width=\textwidth]{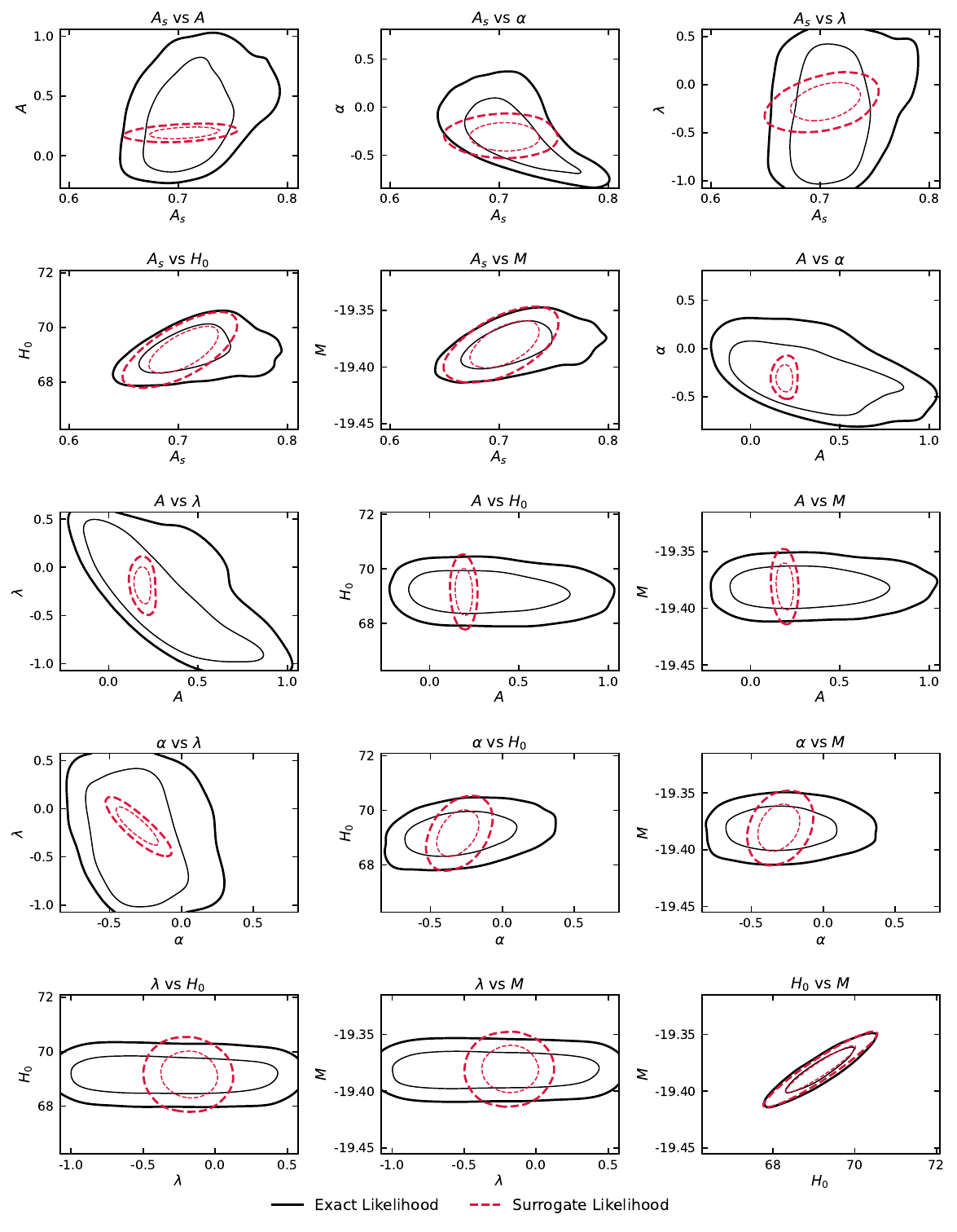}
    \caption{\textbf{Comparison of two-dimensional joint posterior credible regions ($68.3\%$ and $95.4\%$) obtained from the exact likelihood (black solid contours) and the surrogate-assisted MCMC analysis (red dashed contours).} The outcome of the intended dense training is significantly prominent; the $H_0-M$ plots for both methods are in good agreement since they were initially converged and concentrated in the high-likelihood region, while the remaining parameters appear to be highly resolved with associated correlations.}
\label{fig:posterior_overlay_2D}
\end{figure}

To investigate the differences in the posterior distributions, we further compare the results obtained from the exact and surrogate-assisted MCMC analyses. Figure~\ref{fig:posterior_overlay_1D} shows the marginalized one-dimensional overlapping posterior distributions for all six model parameters, and Figure~\ref{fig:posterior_overlay_2D} presents the corresponding pairwise two-dimensional overlapping posterior contours, while Table~\ref{tab:mcmc_results_comparison} represents the normalized shifts in the respective posterior median values.

\begin{table}[htbp]
    \centering
    \caption{\textbf{Comparison of parameter constraints from the exact-likelihood and surrogate-assisted MCMC analyses.}}
    \label{tab:mcmc_results_comparison}
    \vspace{2mm}
    \resizebox{\textwidth}{!}
    {
    \begin{tabular}{|l|c|c|c|c|c|}
        \hline
        & \multicolumn{2}{c|}{\textbf{Exact Likelihood}} & \multicolumn{2}{c|}{\textbf{Surrogate Likelihood}} & \\
        \cline{2-5}
        Parameter & Best-fit & Marginalized ($1\sigma$) & Best-fit & Marginalized ($1\sigma$) & Normalized Shift ($\Delta_{\mathrm{Med}}$) \\
        \hline
        $A_s$     & $0.7126$  & $0.7099_{-0.0249}^{+0.0288}$ & $0.7081$  & $0.7050_{-0.0210}^{+0.0193}$ & $0.15\sigma$ \\[1.2ex]
        $A$       & $0.3966$  & $0.3098_{-0.2625}^{+0.3220}$ & $0.2166$  & $0.1932_{-0.0317}^{+0.0280}$ & $0.40\sigma$ \\[1.2ex]
        $\alpha$  & $-0.3604$ & $-0.2976_{-0.2263}^{+0.2690}$ & $-0.3371$ & $-0.3096_{-0.0876}^{+0.0963}$ & $0.05\sigma$ \\[1.2ex]
        $\lambda$ & $-0.4635$ & $-0.3208_{-0.4524}^{+0.5049}$ & $-0.1937$ & $-0.1813_{-0.1276}^{+0.1205}$ & $0.28\sigma$ \\[1.2ex]
        $H_0$     & $69.20$   & $69.15_{-0.53}^{+0.52}$       & $69.15$   & $69.16_{-0.54}^{+0.54}$       & $0.01\sigma$ \\[1.2ex]
        $M$       & $-19.379$ & $-19.381_{-0.013}^{+0.012}$   & $-19.380$ & $-19.381_{-0.013}^{+0.013}$   & $0.00\sigma$ \\
        \hline
    \end{tabular}
    }
\end{table}

Figure~\ref{fig:posterior_overlay_1D} shows that the baseline expansion parameters $H_0$ and $M$ are essentially unchanged between the two analyses. For $A_s$, the two analyses also identify a very similar preferred region, with the main difference arising due to the broader asymmetric tail of the exact posterior. In contrast, the parameters $A$, $\alpha$, and $\lambda$ exhibit substantial differences in their respective posterior widths. In these cases, the surrogate-assisted posteriors are more concentrated than the corresponding exact-likelihood posteriors, indicating that the surrogate retains well-defined posterior support in these parameter directions despite the presence of parameter degeneracies.

The two-dimensional overlapping posterior contours in Figure~\ref{fig:posterior_overlay_2D} provide a complementary assessment of the parameter degeneracies. The surrogate-assisted contours generally follow the principal orientation of the exact-likelihood contours, indicating that the dominant correlations among the cosmological parameters are retained. This agreement is particularly evident for the correlated combinations involving $A_s$--$H_0$, $A_s$--$M$, $H_0$--$M$, and several of the $\alpha$- and $\lambda$-related parameter pairs. However, the surrogate-assisted contours are generally narrower than the exact contours for several combinations, consistent with the differences observed in the one-dimensional marginalized posteriors. The overlay plots show that the posterior plots for well-constrained parameters in the conventional exact-likelihood analysis are in close agreement with the corresponding ones in the surrogate analysis. For the poorly constrained parameters, the surrogate provides a more concentrated posterior representation in the region of high likelihood.

Lastly, Table~\ref{tab:mcmc_results_comparison} summarises the quantitative comparison between the parameter constraints derived from the exact likelihood MCMC analysis and the surrogate-assisted framework, along with their corresponding normalized median shifts, given as $\Delta_{\mathrm{Med}} \equiv D_\sigma(\theta_{\mathrm{exact}}, \theta_{\mathrm{surr}})$ (Eq.~\eqref{eq:unified_shift_metric}). It demonstrates that the surrogate-assisted analysis accurately reproduces the median values of the marginalized posteriors of the exact probability, with all the median shifts being less than $0.5\sigma$. The agreement is particularly strong for $\alpha$, $H_0$, and $M$, with shifts of $0.05\sigma$, $0.01\sigma$, and $0.00\sigma$, respectively. The largest shift occurs for $A$ ($0.40\sigma$), while $A_s$ and $\lambda$ remain within $0.3\sigma$. 

These comparisons demonstrate that the surrogate likelihood captures the preferred location of the exact posterior, while some loss of posterior width remains for parameters whose exact likelihood exhibits broader or more non-Gaussian structure. This provides an important validation of the surrogate beyond pointwise $\chi^2$ prediction accuracy, since the comparison directly tests the propagation of surrogate errors into the Bayesian parameter inference. The overall consistent agreement between the posterior constraints supports the use of the machine learning surrogate as a computationally efficient approach for posterior exploration in the presence of parameter degeneracies and long autocorrelations.

\section{Results and Discussion}
\label{sec:Results}

In this section, we study the cosmological implications of the parameter constraints of the Finslerian--MCG model inferred from the joint analysis of $1700$ total observational data points from CC, Pantheon$+$SH$0$ES and DESI DR$2$. We use the surrogate-derived median posterior values with $1\sigma$ credible intervals for physical interpretation and model comparison. Finally, we compare the marginalized parameter constraints of the modified framework with the standard $\Lambda$CDM model. 

\subsection{Cosmological Interpretation of the Posterior Constraints}

The joint analysis constrains the present-day expansion rate and the absolute magnitude of SNe~Ia to $H_0 = 69.16 \pm 0.54~\mathrm{km\,s^{-1}\, Mpc^{-1}}$ and $M = -19.381 \pm 0.013$, respectively. This inferred $H_0$ lies intermediate between early-universe and local distance-ladder calibrations. The median value of $H_0$ exhibits a $2.36\sigma$ tension relative to the Planck~$2018$ result \citep{Planck2020Cosmology}, while the tension with the local SH$0$ES distance-ladder measurement \citep{Riess2022SH0ES} is approximately $3.31\sigma$.

Within the Modified Chaplygin Gas framework, the parameter $A_s = 0.7050_{-0.0210}^{+0.0193}$ characterizes the relative contribution of the late-time dark-energy-like component in the normalized cosmic expansion history. Although $A_s$ and the standard $\Lambda$CDM dark-energy density parameter $\Omega_\Lambda$ arise from different cosmological parameterizations, their numerical values can be compared as effective measures of the late-time dark-energy-like contribution. The inferred $A_s$ is in close statistical agreement with the Planck $2018$ baseline value $\Omega_\Lambda = 0.6847\pm 0.0073$ \citep{Planck2020Cosmology}, with a difference of approximately $0.94\sigma$.  The linear equation-of-state parameter is constrained to $A = 0.1932_{-0.0317}^{+0.0280}$.

To understand the physical behaviour of the fluid, consider the effective equation-of-state (EoS) parameter \citep{Copeland2006Dynamics},
\begin{equation}
    w_{\rm MCG} \equiv \frac{p_{\rm MCG}}{\rho} = A - \frac{B}{\rho^{1+\alpha}}.
\end{equation}
The posterior yields $\alpha = -0.3096_{-0.0876}^{+0.0963}$, placing the $68.3\%$ credible interval entirely within the range $-1 < \alpha < 0$ and giving $1+\alpha = 0.6904 > 0$ \citep{Li2019MCG}. In the dense early-universe regime, the non-linear Chaplygin contribution to the effective EoS, $\frac{B}{\rho^{1+\alpha}}$, becomes negligible relative to the linear $A$ term, leading to
\begin{equation}
    w_{\rm MCG} \rightarrow A \simeq 0.1932.
\end{equation}
Thus, in the asymptotically high-density limit, the MCG component approaches a positive-pressure state rather than the pressureless-dust limit. The negative posterior value for $\alpha$ corresponds to a sublinear density dependence of the Chaplygin pressure contribution, $p_{\rm Ch} \propto -\rho^{-\alpha} = -\rho^{0.3096}$, while its contribution relative to the energy density scales as
\begin{equation}
    \frac{p_{\rm Ch}}{\rho} \propto -\rho^{-(1+\alpha)} = -\rho^{-0.6904}.
\end{equation}
As cosmic expansion dilutes the energy density, this negative-pressure term gradually takes over, driving the effective EoS toward negative values.

Looking toward the distant future, the late-time asymptotic behaviour can be obtained from the Finslerian--MCG density evolution. Defining the background geometric parameter
\begin{equation}
    D \equiv 3(1+A) + \lambda\left(1 + \frac{3A}{2}\right),
\end{equation}
the posterior medians give $D \simeq 3.3458$. From  Eq~\ref{eq:A_s}
\begin{equation}
    A_s = \frac{3\left(1 + \frac{\lambda}{2}\right)B}{D\rho_0^{1+\alpha}},
\end{equation}
the energy density approaches a finite, non-zero asymptotic value, $\rho_\infty = \rho_0 A_s^{1/(1+\alpha)}$ (from Eq~\ref{eq:rho_a} and Eq~\ref{eq:A_s}). The corresponding asymptotic EoS is
\begin{equation}
    w_\infty = A - \frac{D}{3\left(1 + \frac{\lambda}{2}\right)},
\end{equation}
which evaluates to $w_\infty \simeq -1.033$. While the present-day fluid remains non-phantom, with $w_0 \equiv A - B/\rho_0^{1+\alpha} \simeq -0.671$, the effective MCG equation of state therefore evolves from a non-phantom present value towards a mildly phantom asymptotic regime \citep{Caldwell2002Phantom}. This late-time behaviour arises from the combined effect of the MCG pressure contribution and the modified Finslerian background dynamics, rather than from the value of $\alpha$ alone.

The Finslerian anisotropy parameter is constrained to $\lambda = -0.1813_{-0.1276}^{+0.1205}$, favoring a mild negative geometric deformation within the weak-anisotropy regime \citep{Chakraborty2026FinslerRanders}. The posterior median is statistically consistent with the isotropic baseline ($\lambda=0$) at the $1.46\sigma$ level, while showing good agreement with $\lambda=-0.06 \pm 0.048$ reported in \citep{Chakraborty2026FinslerRanders} at the $0.91\sigma$ level. This negative geometric contribution reduces the background parameter $D$ from its isotropic value, $D_{\lambda=0} = 3(1+A) = 3.5796$ to $D \simeq 3.3458$. Consequently, the effective redshift exponent governing the density evolution, $n = (1+\alpha)D$, decreases from $n_{\lambda=0} \simeq 2.4714$ to $n \simeq 2.3099$. Within the present framework, the negative Finslerian deformation reduces this dilution exponent, leading to a slower dilution of the energy density than in the standard isotropic limit. 

Moreover, the present cosmological model has also been compared to the standard $\Lambda\mathrm{CDM}$ model using the same configuration of observational datasets to observe the difference between the baseline parameters ($H_0$, $M$ and $A_s$) and the anisotropy parameter ($\lambda$). Table~\ref{tab:parameter_constraints} summarizes the comparison between the marginalized posterior constraints of the present Finslerian--MCG model and the standard $\Lambda\mathrm{CDM}$ model. The statistical difference in units of combined standard deviation is evaluated using the same metric, $\text{Difference } (\sigma) \equiv D_\sigma(\theta_{\mathrm{F}}, \theta_{\Lambda\mathrm{CDM}})$ (Eq.~\ref{eq:unified_shift_metric}). The present modified framework is observed to exhibit a much smaller difference in $H_0$ and $A_s$ (eff. $\Omega_{\Lambda}$) as $0.06\sigma$ and $0.07\sigma$, respectively. The difference for $M$ also remains less than $0.3\sigma$. But for the Finslerian anisotropy parameter, $\lambda$, the modified framework exhibits a difference of $1.46\sigma$, indicating a mild departure from the standard isotropic limit.

\begin{table}[htbp]
\centering
    \caption{\textbf{Comparison of marginalized posterior constraints for the Finslerian--MCG model against the standard $\Lambda$CDM model using the joint CC $+$ Pantheon$+$SH0ES $+$ DESI DR$2$ dataset.}}
    \label{tab:parameter_constraints}
    \vspace{2mm}
    \begin{tabular}{|l|c|c|c|}
        \hline
        Parameter & Finslerian--MCG & $\Lambda\mathrm{CDM}$ & Difference ($\sigma$) \\
        \hline
        $H_0$ & $69.16 \pm 0.54$ & $69.12 \pm 0.42$ & $0.06\sigma$ \\[1.2ex]
        $M$ & $-19.381 \pm 0.013$ & $-19.385 \pm 0.011$ & $0.23\sigma$ \\[1.2ex]
        $A_s$ (eff.\ $\Omega_\Lambda$) & $0.7050_{-0.0210}^{+0.0193}$ & $0.7065 \pm 0.0068$ & $0.07\sigma$ \\[1.2ex]
        $\lambda$ & $-0.1813_{-0.1276}^{+0.1205}$ & $0\text{ (fixed)}$ & $1.46\sigma$ \\[1.2ex]
        $A$ & $0.1932_{-0.0317}^{+0.0280}$ & --- & --- \\[1.2ex]
        $\alpha$ & $-0.3096_{-0.0876}^{+0.0963}$ & --- & --- \\
        \hline
    \end{tabular}
\end{table}

\subsection{Model Selection and Statistical Comparison (AIC and BIC)}

Moreover, we evaluate the quantitative statistical performance of the present cosmological model against the standard $\Lambda$CDM model with their respective number of free parameters. The comparison is performed using the minimum chi-squared statistic $\chi^2_{\min}$ together with the Akaike Information Criterion (AIC)  \citep{Akaike1974} and the Bayesian Information Criterion (BIC) \citep{Schwarz1978}. These are the standard statistical measures that simultaneously account for both the goodness of fit and the complexity of the model. For a particular model with $p$ free parameters studied using the $N$ ($1700$) observational data points, the criteria are defined as \citep{Schwarz1978, Akaike1974, Liddle2004}

\begin{equation}
    \mathrm{AIC} = \chi^2_{\min} + 2p, \quad \mathrm{BIC} = \chi^2_{\min} + p \ln N.
\end{equation}

The performance of the model relative to the flat $\Lambda$CDM is compared using information differences.
\begin{equation}
    \Delta\mathrm{AIC} = \mathrm{AIC}_{\mathrm{model}} - \mathrm{AIC}_{\Lambda\mathrm{CDM}}, \quad 
    \Delta\mathrm{BIC} = \mathrm{BIC}_{\mathrm{model}} - \mathrm{BIC}_{\Lambda\mathrm{CDM}}.
    \label{eq:Delta}
\end{equation}

We follow the standard information theoretic framework of \citep{Burnham2002}, which states that an information difference of $|\Delta\mathrm{AIC}| \le 2$ indicates that the models are statistically comparable to the baseline. Values in the range $2 < |\Delta\mathrm{AIC}| \le 4$ indicate some evidence against the model with higher AIC, while $4 < |\Delta\mathrm{AIC}| \le 7$ corresponds to considerably less statistical support for the model with higher AIC, whereas $|\Delta\mathrm{AIC}| > 10$ signifies essentially no statistical support for the model with higher AIC. 

For the Bayesian Information Criterion, the article adopts the scale \citep{Kass1995}, where $|\Delta\mathrm{BIC}| < 2$ indicates weak or negligible evidence for a difference between the models, $2 \le |\Delta\mathrm{BIC}| < 6$ represents positive evidence for the model with lower BIC and $6 \le |\Delta\mathrm{BIC}| < 10$ indicates strong evidence while $|\Delta\mathrm{BIC}| \ge 10$ provides very strong evidence in favour of the model with lower BIC.

\begin{table}[htbp]
    \centering
    \caption{\textbf{Statistical Comparison of the Finslerian–MCG model and the standard $\Lambda$CDM model using the joint CC $+$ Pantheon$+$SH$0$ES $+$ DESI DR$2$ dataset ($N = 1700$).}}
    \label{tab:aic_bic_combination}
    \vspace{2mm}
    \begin{tabular}{|l|c|c|c|c|c|c|c|}
        \hline
        Model & $p$ & $\chi^2_{\min}$ &  AIC & BIC & $\Delta\mathrm{AIC}$ & $\Delta\mathrm{BIC}$ \\
        \hline
        $\Lambda\mathrm{CDM}$ & $3$ & $1503.115$ & $1509.115$ & $1525.430$ & $0.000$ & $0.000$ \\
        Finslerian--MCG & $6$ & $1501.400$ & $1513.400$ & $1546.030$ & $+4.285$ & $+20.600$ \\
        \hline
    \end{tabular}
\end{table}

Table~\ref{tab:aic_bic_combination} clearly demonstrates that the present Finslerian modified Chaplygin gas model provides a better fit to the joint dataset combination than the standard $\Lambda$CDM model, yielding a lower minimum chi-squared with $\Delta\chi^2 = -1.715$. However, when accounting for model complexity, the three additional free parameters ($\Delta p = 3$) introduce penalty costs of $2\Delta p = 6$ to AIC and $\Delta p \ln N = 22.315$ to BIC. This results in $\Delta\mathrm{AIC} = +4.285$, demonstrating that while the Finsler--MCG model remains competitive, it receives substantially less statistical support compared to the baseline $\Lambda$CDM model. Furthermore, $\Delta\mathrm{BIC} = +20.600$ heavily penalizes the extra parameter freedom due to the large dataset size ($N = 1700$), providing very strong evidence in favour of standard flat $\Lambda$CDM as the more economical description of the background cosmic expansion.

\section{Conclusion}
\label{sec:Conclusion}
In this paper, we formulated a cosmological model that incorporates anisotropy arising from intrinsic geometric modifications, using the Finslerian framework and the Modified Chaplygin Gas as a unified description of the dark sector. The anisotropy has been assumed to evolve linearly in the modified Friedmann equation, which was used to derive the expansion history $H(z)$. The model was then studied observationally using recent datasets to derive parameter constraints and assess how well these datasets align with the cosmological model. The conventional method used for constraining the parameters was found to struggle with the poor mixing of parameter chains and long autocorrelation times due to the model's intrinsic parameter degeneracies, resulting in a requirement of increased computational budget.

To reduce computational expense, we employed a machine-learning-assisted algorithm, Heteroscedastic Deep Ensemble with MC Dropout, incorporating a distance- and $\Delta\chi^2$-gated local residual corrector, to accelerate parameter sampling and explore the posterior structure in the high-likelihood region. This likelihood emulator is intended to provide a computationally efficient approximation of the likelihood in the posterior dominant region. We tested the algorithm for validity and ran several diagnostics, including $\mathrm{MAE}$, $\mathrm{RMSE}$, and $\mathrm{R}^2$, yielding $0.4170$, $0.5770$, and $0.9983$, respectively, in the posterior-relevant region. The value $\mathrm{R}^2$ near unity indicates strong predictive accuracy in the high-likelihood region. The emulator incorporated MC Dropout, which accounts for the predictive uncertainty in addition to predicting $\chi^2$. 

This surrogate replaced the expensive exact evaluation of the $\chi^2$ with the fast, deterministic predictive $\chi^2$ in the same conventional parameter sampling algorithm. Both sampling procedures were carried out using the same configuration of iterations and walkers. While the exact evaluation of $\chi^2$ resulted in unconverged sampling, the surrogate-predicted $\chi^2$ resulted in converged sampling verified against the convergence diagnostics. The minimum $\chi^2$ was recalculated using the exact likelihood evaluation for the surrogate-assisted analysis (yielding, $1501.3998$) and was found to be in close agreement with the one obtained from exact likelihood in conventional analysis (yielding, $1501.40$) corresponding to a difference of $\Delta \chi^2=0.0002$, while the predicted surrogate-derived $\chi^2_\mathrm{min}$ (yielding, $1501.5215 \pm 0.2052$) differed by $0.1217$ from the exact reevaluation for the same best-fit point.

The results were examined to identify whether the ML surrogate introduces any artificial modification to the correlated posterior structure using the overlapping $1$D and $2$D posterior plots and the normalized shift in their respective median values. The overlapping posterior plots were found to be in close agreement with each other for the parameters that were well constrained conventionally. However, for parameters that did not converge within the given computational budget, the overlay plots showed a substantial difference, providing tighter constraints on the parameters along with the identification of correlation directions. The normalized shifts in the median values for all the parameters were found to be less than $0.5\sigma$. The maximum shift of $0.40\sigma$ was observed for $A$, which exhibited the maximum autocorrelation length ($\tau = 1578.3$) and the maximum Gelman-Rubin statistic ($\hat{R}=1.096$), while the minimum shift of $0.00\sigma$ was observed for $M$ which exhibited the minimum autocorrelation length ($\tau = 449.6$) and the minimum Gelman-Rubin statistic ($\hat{R}=1.015$) in the exact likelihood analysis. Thus, the machine learning surrogate did not introduce a significant artificial modification to the sampling analysis within the diagnostics performed and can be taken as an efficient alternative for the Bayesian analysis of the cosmological model, which struggles due to the model's intrinsic parameter degeneracies. The observationally induced $H_0-M$ correlation remained unaltered and closely consistent between the exact likelihood and surrogate-assisted analyses, indicating that the surrogate preserves the degeneracy associated with the observational parameters. 

The surrogate-derived best-fit points were used to study the cosmological implications of the parameter constraints. The best-fit values for the model parameters $\{A_s, A, \alpha, \lambda, H_0, M\}$ were obtained as $\{0.7081, 0.2166, -0.3371, -0.1937, 69.15, -19.380\}$. These constraints provide a consistent description of the late-time expansion history, with a small non-zero Finslerian anisotropy parameter, indicating a mild departure from the isotropic limit and MCG parameters that indicate a departure from the simplest barotropic fluid description.   

The derived expansion history model was further assessed for statistical evidence against the standard $\Lambda$CDM model. The model was found to exhibit a slightly better fit in terms of minimum $\chi^2$ to the joint analysis. The extended model contains additional free parameters, and therefore the corresponding AIC and BIC values explicitly account for this increase in model complexity. The AIC and BIC values demonstrate the positive statistical support and a very strong preference for the simpler $\Lambda$CDM model, respectively, indicating that the improvement in the minimum $\chi^2$ is not sufficient to justify the additional parameters of the extended model.  

The present analysis provides observational constraints on the Finslerian--MCG model using background expansion data. Looking ahead, high-precision observations of cosmic structure growth and independent anisotropy constraints could potentially further constrain the MCG and Finslerian parameters, which may help assess whether the non-Riemannian geometrical modification considered here can be distinguished from the standard $\Lambda$CDM dark-energy framework. These observations may also help identify any significant departure from the standard Riemannian framework in describing the observed cosmic acceleration. Building on the methodological framework established in this paper, more advanced machine learning architectures can be developed and extended to a larger parameter space and additional cosmological datasets for accelerating Bayesian sampling while maintaining statistical precision and the accuracy of the inferred posterior distributions.

\section*{Data availability}

The citations to datasets utilized are provided in the paper.

\section*{Declaration of competing interest}

The authors declare that they have no known competing financial interests or personal relationships that could have appeared to influence the work reported in this paper.

\section*{Acknowledgements}

S. Seth and R. Chaubey both acknowledge SERB, New Delhi, for the financial assistance through project No. CRG/2023/004560 (P-07/1328).

\bibliographystyle{mnras}
\bibliography{ref}

\end{document}